\pdfoutput=1
\documentclass[preprint]{aastex63}

\newcommand{\HI}{H$\,${\sc i}}

\newcommand{\MassSDSF}{$\Sigma_{*\scriptsize{\textnormal{,SF}}}$}
\newcommand{\MassSDEnv}{$\Sigma_{*\scriptsize{\textnormal{,env}}}$}
\newcommand{\HISDSF}{$\Sigma_{\scriptsize{\textnormal{H}\,\textnormal{\sc{i}}\,\textnormal{,SF}}}$}
\newcommand{\HISDenv}{$\Sigma_{\scriptsize{\textnormal{H}\,\textnormal{\sc{i}}\,\textnormal{,env}}}$}
\newcommand{\HISDrad}{$\Sigma_{\scriptsize{\textnormal{H}\,\textnormal{\sc{i}}\,\textnormal{,rad}}}$}
\newcommand{\HISDdiff}{$\Sigma_{\scriptsize{\textnormal{H}\,\textnormal{\sc{i},SF}}} - \Sigma_{\scriptsize{\textnormal{H}\,\textnormal{\sc{i}}\,\textnormal{,rad}}}$}
\newcommand{\MCO}{$M_{\textnormal{\scriptsize{CO}},vir}$}
\newcommand{\COSD}{$\Sigma_{\scriptsize{\textnormal{CO}}}$}
\newcommand{\darkgas}{$M_{\scriptsize{Dark \textrm{ H}_2}}$/$M_{\scriptsize{MC}}$}
\newcommand{\HISD}{$\Sigma_{\scriptsize{\textnormal{H}\,\textnormal{\sc{i}}}}$}

\received{October 7, 2021}
\revised{January 10, 2022}
\accepted{January 23, 2022}
\submitjournal{\aj}

\shorttitle{WLM CO Environments}
\shortauthors{Archer et al.}
\graphicspath{{./}{figures/}}

\begin{document}

\title{The Environments of CO Cores and Star Formation in the Dwarf Irregular Galaxy WLM}

\correspondingauthor{Haylee Archer}
\email{harcher@lowell.edu}

\author[0000-0002-8449-4815]{Haylee N. Archer}
\affiliation{School of Earth and Space Exploration \\
Arizona State University \\
Tempe, AZ 85287 USA}
\affiliation{Lowell Observatory \\
1400 W Mars Hill Rd \\
Flagstaff, AZ 86001 USA}

\author[0000-0002-3322-9798]{Deidre A. Hunter}
\affiliation{Lowell Observatory \\
1400 W Mars Hill Rd \\
Flagstaff, AZ 86001 USA}

\author[0000-0002-1723-6330]{Bruce G.\ Elmegreen}
\affiliation{IBM T.\ J.\ Watson Research Center \\
1101 Kitchawan Road \\
Yorktown Heights, NY 10598 USA}

\author[0000-0002-8736-2463]{Phil Cigan}
\affiliation{George Mason University \\
4400 University Dr. \\
Fairfax, VA 22030-4444, USA}

\author[0000-0003-1268-5230]{Rolf A. Jansen}
\affiliation{School of Earth and Space Exploration \\
Arizona State University \\
Tempe, AZ 85287 USA}

\author[0000-0001-8156-6281]{Rogier A. Windhorst}
\affiliation{School of Earth and Space Exploration \\
Arizona State University \\
Tempe, AZ 85287 USA}

\author[0000-0001-9162-2371]{Leslie K. Hunt}
\affiliation{INAF, Osservatorio Astrofisico di Arcetri \\
Largo E Fermi 5 \\
50125 Firenze Italy}

\author[0000-0002-5307-5941]{Monica Rubio}
\affiliation{Departamento de Astronom\'{i}a \\
Universidad de Chile \\
Casilla 36-D, 8320000 Santiago, Chile}



\begin{abstract}
The low metallicities of dwarf irregular galaxies (dIrr) greatly influence the formation and structure of molecular clouds. These clouds, which consist primarily of H$_2$, are typically traced by CO, but low metallicity galaxies are found to have little CO despite ongoing star formation.
In order to probe the conditions necessary for CO core formation in dwarf galaxies, we have used the catalog of Rubio et al.\ (2022, in preparation) for CO cores in WLM, a Local Group dwarf with an oxygen abundance that is 13\% of solar. 
Here we aim to characterize the galactic environments in which these 57 CO cores formed. We grouped the cores together based on proximity to each other and strong FUV emission, examining properties of the star forming region enveloping the cores and the surrounding environment where the cores formed. We find that high \HI\ surface density does not necessarily correspond to higher total CO mass, but regions with higher CO mass have higher \HI\ surface densities. We also find the cores in star forming regions spanning a wide range of ages show no correlation between age and CO core mass, suggesting that the small size of the cores is not due to fragmentation of the clouds with age. The presence of CO cores in a variety of different local environments, along with the similar properties between star forming regions with and without CO cores, leads us to conclude that there are no obvious environmental characteristics that drive the formation of these CO cores.
\end{abstract}



\section{Introduction}\label{sec:intro}
Wolf-Lundmark-Melotte (WLM) is a Local Group, dwarf irregular (dIrr) galaxy at a distance of $985\pm33$ kiloparsecs (kpc) \citep{Leaman_2012}. Like other dwarf galaxies, the mass and metallicity of WLM are low, with a total stellar mass of $1.62 \times 10^7 M_\odot$ \citep{Zhang_2012} and metallicity of 12$+$log(O/H)$=$7.8 \citep{Lee_2005}. WLM is an isolated galaxy, and the large spatial distances between it and both the Milky Way and M31 indicate a low probability of past interaction with either \citep{Teyssier_2012, Albers_2019}. The low mass, low metallicity, distance, and isolation of WLM make it an ideal laboratory for understanding star formation in undisturbed dwarf galaxies.

Star formation in galaxies is believed to be mostly regulated by molecular gas found in giant molecular clouds (GMCs) in the interstellar medium (ISM) \citep{Kennicutt_Jr__1998, McKee_2007}. The most abundant species in these molecular clouds is molecular hydrogen (H$_2$), which is nearly impossible to observe in the typical conditions of the cold ISM because it does not possess a permanent dipole moment and thus no dipolar rotational transitions \citep{Bolatto_2013}. As such, H$_2$ is traced using indirect methods, the most common of which is through the measurement of low rotational lines of carbon monoxide (CO). Despite being much less abundant than H$_2$ in molecular clouds, CO is easily excited even in the cold ISM.

Many low-metallicity dwarf galaxies are found to have little CO despite ongoing star formation \citep{Elmegreen_1980_2}, which disputes the standard model of star formation in CO-rich molecular clouds. If the small amount of detected CO is translated to the total H$_2$ of the cloud using the standard conversion factor, $X_{CO}$,
of more massive galaxies, the high inferred star formation efficiency of the dwarfs would make them outliers on the Schmidt-Kennicutt relation \citep{Kennicutt_Jr__1998, Madden_2018}. \citet{Elmegreen_1989} finds that the increase in CO formation time at lower metallicity could result in the disruption and dissociation of H$_2$ before CO can form anywhere but in the cores of larger clouds. This longer CO formation time is partly because lower metallicity also corresponds to lower dust abundance, which allows far-ultraviolet (FUV) photons to photodissociate CO molecules in the molecular cloud and leave behind smaller CO cores \citep{Elmegreen_1980, Taylor_1998, Schruba_2012}. H$_2$ is self-shielded from the FUV photons and can survive in the photodissociation region (PDR). This H$_2$ gas that is not traced by the CO cores is referred to as ``dark" gas \citep{Wolfire_2010}. There is strong evidence that the observed lack of CO at low metallicities is a natural consequence of the lower carbon and oxygen abundances as metallicity decreases, with the result that the H$_2$ is primarily associated with the so-called CO-dark molecular gas \citep{Wolfire_2010, 2011, Pineda_2014, Cormier_2017}.

\begin{figure*}[htb!]
\epsscale{0.7}\plotone{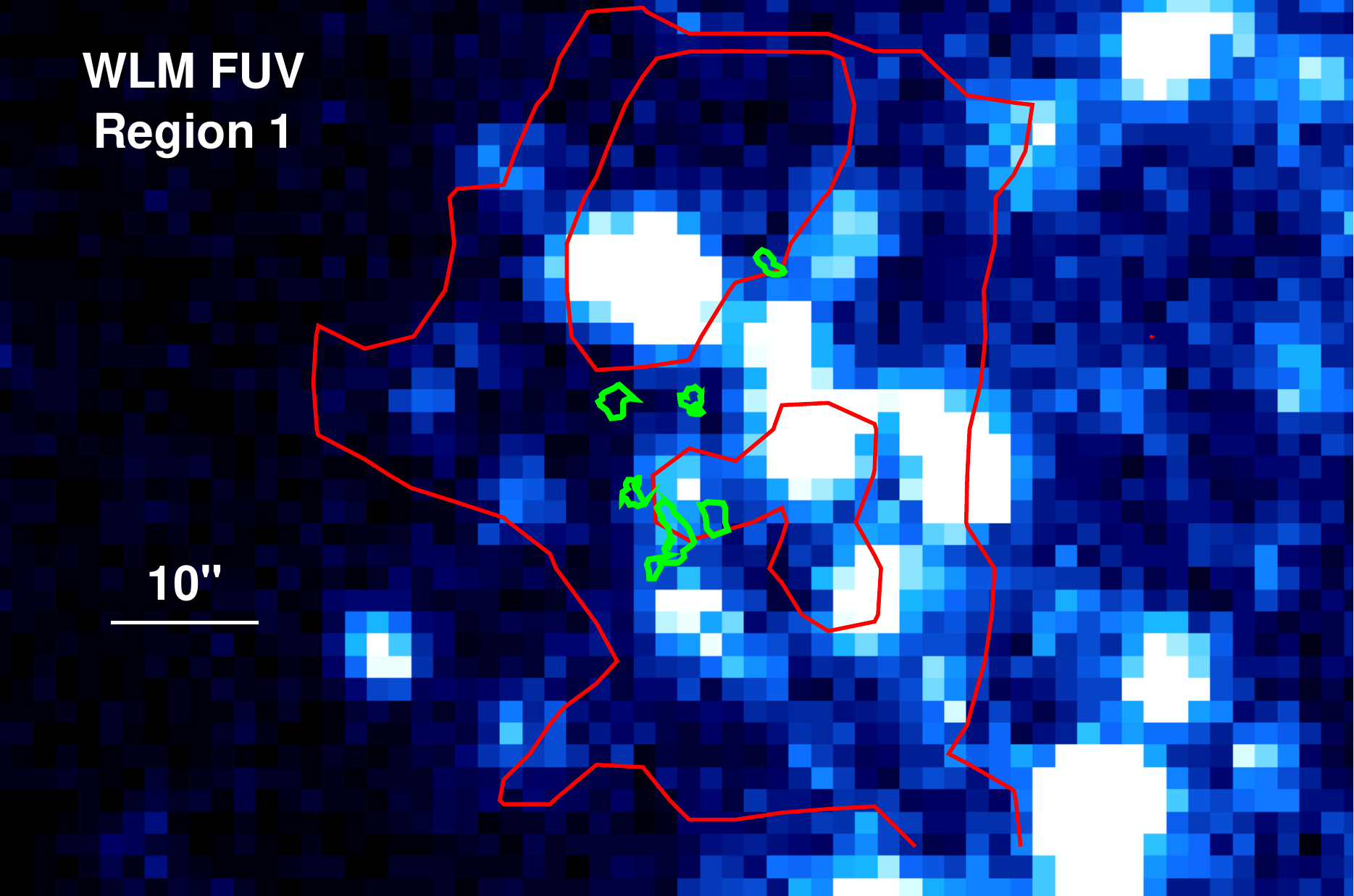}
\caption{FUV image of a WLM star forming region (region 1 in this paper, as in Figures \ref{fig:fuvWreg}--\ref{fig:presWreg}) overlaid with the PDR indicated by [C$\,${\sc ii}]$\lambda$158 $\mu$m contours (red) of $2.5 \times 10^{-19}$ and $4.6 \times 10^{-19}$ W m$^{-2}$ pix$^{-1}$, for pixels of 3\farcs13 per side, from \citet{Cigan_2016} and 6 CO core contours (green) from \citet{Rubio:2015}. 
\label{fig:reg1Contour}}
\end{figure*}

Following the discovery by \citet{Elmegreen_2013} of CO(3--2) in two star forming regions of WLM using the APEX telescope, \citet{Rubio:2015} used pointed CO(1--0) of these regions with the the Atacama Large Millimeter Array (ALMA) to map 10 CO cores for the first time at an oxygen abundance that is 13\% of the solar value \citep{Lee_2005, Asplund_2009}. The PDR region as traced by the C$\,${\sc ii} observations surrounding six of the discovered cores is five times wider than the cluster of cores. This indicates that molecular cloud structure at lower metallicities consists of thicker H$_2$ shells and smaller CO cores compared to those seen in the Milky Way \citep{Rubio:2015,Cigan_2016}. An FUV image of the region with that PDR and the six detected CO cores overlaid as contours is shown in Figure \ref{fig:reg1Contour}. Rubio et al.\ (2022, in preparation) has since mapped most of the star forming area of WLM with pointed ALMA CO(2--1) observations and detected an additional 47 cores.

This paper seeks to characterize the galactic environments in which these star forming CO cores formed in WLM to determine (1) if the CO cores have the same properties in different local environments, (2) if areas where CO has formed have different properties from star forming regions without detected CO, (3) the nature of the stellar populations surrounding the molecular clouds, and (4) the relationship between CO and star formation. The paper is organized as follows. In Section \ref{sec:data} we introduce our multi-wavelength data and describe our region selection and definitions of their environment, along with our methods for determining the region age, stellar mass surface density, and CO-dark gas. We present our results in Section \ref{sec:results} and discuss our findings in Section \ref{sec:discussion}.

\begin{figure*}[htb!]
\epsscale{1.1}\plotone{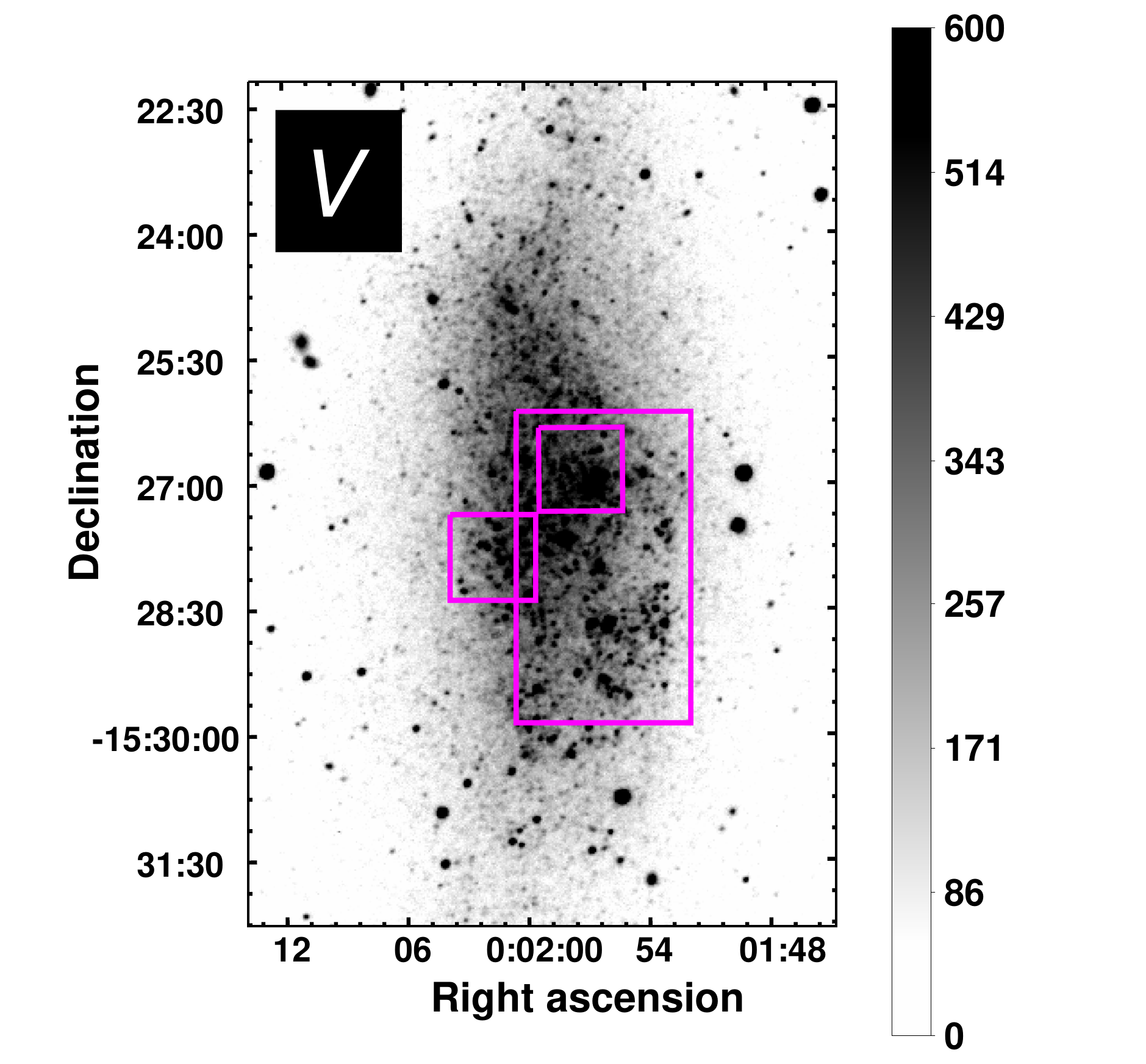}
\caption{$V$ band image of WLM showing the ALMA field of view (FOV) from \citet{Rubio:2015} (smaller magenta squares) and Rubio et al.\ (2022, in preparation) (larger magenta rectangle). The orientation of the image is such that North is up and East is to the left. Colorbar values can be converted from counts in one pixel (DN/pixel) to Johnson magnitudes with the equation: 
$V_{Johnson} = 0.0157(B-V)_{Johnson} -2.5\log V_{counts} +29.41$.
\label{fig:v_fov}}
\vspace{-1.5em}
\end{figure*}

\section{Data} \label{sec:data}
Two star-forming regions of WLM were imaged in CO(1--0) with ALMA in Cycle 1 by \citet{Rubio:2015} where 10 CO cores were detected with an average radius of 2 pc and average virial mass of $2 \times 10^3$ $M_{\odot}$. Another 47 CO cores were discovered in WLM from Cycle 6 ALMA CO(2--1) observations over much of the star-forming area of the galaxy, which included one of the two regions observed in Cycle 1 (Rubio et al., 2022, in preparation). The beam size of these observations were $0\farcs6 \times 0\farcs5$.
These two resulting catalogues provide characteristics of the CO cores including locations, virial masses, and surfaces densities. We use the sum of the virial masses of the individual CO cores for each region (\MCO) and the median surface density of the individual CO cores in each region (\COSD) to examine any relationships with other star forming and environmental properties.
Figure \ref{fig:v_fov} shows the $V$ band image of WLM overlaid with an outline of the total field of view observed by \citet{Rubio:2015} and Rubio et al.\ (2022, in preparation).

\textit{UBV} images of WLM came from observations using the Lowell Observatory Hall 1.07-m Telescope. Further information on the acquisition and reduction of these images are described by \citet{Hunter_2006}. \HI\ surface density (\HISD) maps and \HI\ surface density radial profiles were acquired by \citet{Hunter_2012} with the Very Large Array (VLA\footnote[1]{The VLA, now the Karl G.\ Jansky Very Large Array, is a facility of the National Radio Astronomy Observatory. The National Radio Astronomy Observatory is a facility of the National Science Foundation operated under cooperative agreement by Associated Universities, Inc. Observations were made during the transition from the Very Large Array to the Karl G.\ Jansky Very Large Array.}) for the Local Irregulars That Trace Luminosity Extremes, The \HI\ Nearby Galaxy Survey (LITTLE THINGS), a multi-wavelength survey of 37 nearby dIrr galaxies and 4 nearby Blue Compact Dwarf (BCD) galaxies. The authors created robust-weighted and natural-weighted \HI\ maps, and we chose to use the robust-weighted maps due to the higher resolution ($6\arcsec$). The  FUV and near-ultraviolet (NUV) images came from the NASA Galaxy Evolution Explorer (GALEX\footnote[2]{GALEX was operated for NASA by the California Institute of Technology under NASA contract NAS5-98034.}) satellite \citep{galex} GR4/5 pipeline, and were further reduced by \citet{Zhang_2012}. We also used stellar mass surface density ($\Sigma_*$) and pressure maps created by \citet{Hunter_2018}. The stellar mass surface density image was determined on a pixel-by-pixel basis based on $B-V$ \citep{Herrmann_2016}, and the pressure map was calculated with the equation
\begin{equation}
\textnormal{P} = 2.934 \times 10^{-55} \times \Sigma_{gas}(\Sigma_{gas} + (\frac{\sigma_{gas}}{\sigma_*})\Sigma_*) \textnormal{ [g/(s$^2$ cm)]}
\end{equation}
where $\Sigma$ is a surface density and $\sigma$ is a velocity dispersion \citep{Elmegreen_1989}. The $\Sigma_{gas}$ in the pressure map comes from the robust-weighted \HI\ map from \citet{Hunter_2012}. Further details on the creation of these images are described by \citet{Hunter_2018}. To gain insight into the formation of the CO cores, we used the pressure, \HI\ surface density, and stellar mass surface density data to characterize the regions within which the
CO cores formed and the environment surrounding the regions. We also used the \textit{UBV}, FUV, and NUV data for determining ages of the young stars in the star-forming regions of the CO cores. 

\subsection{Regions}\label{regmeth}
We grouped the CO cores into regions based on apparent proximity to FUV knots and each other. The size and clustering of region 1 was chosen using both the ALMA mapped CO cores of 
\citet{Rubio:2015} and their [{C$\,${\sc ii}}]$\lambda$158 micron image from the Herschel Space Observatory indicating the PDR that surrounds those cores (Figure \ref{fig:reg1Contour}). The other regions were chosen by eye based on the following criteria: 1) apparent distance to nearby FUV emission within the plane of the galaxy and 2) apparent distance to other CO cores, with the size of the region determined by grouping CO cores that appeared to be closest to the same FUV knots. We then used \textsc{SExtractor} \citep{Bertin_1996} with a detection and analysis threshold of 3 sigma, a minimum of 10 pixels above the threshold for detection, and a background mesh and filter size of 50 and 7 respectively to objectively identify the brighter FUV knots. We then computed the distances between the center of the CO cores and the center of the detected FUV sources to determine how far each CO core in a region is from its closest detected FUV knot.

\begin{figure*}[b!]
\epsscale{1.1}\plotone{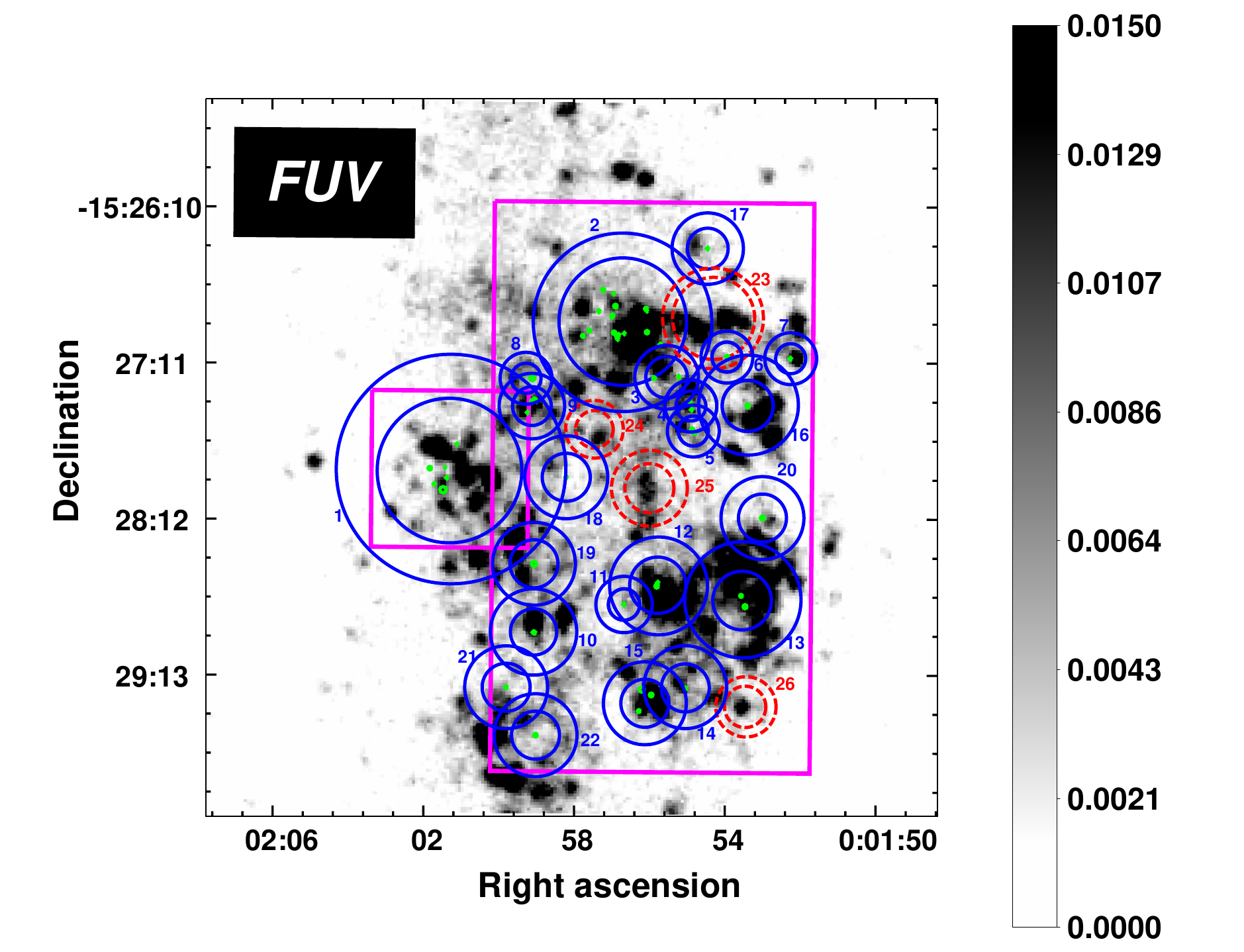}
\caption{FUV image of WLM showing the ALMA FOV of region 1 from \citet{Rubio:2015} (smaller magenta square) and of the Rubio et al.\ (2022, in preparation) survey (larger magenta rectangle), regions and annuli defined here (blue circles for regions with CO cores and red dashed circles for the regions without CO cores), and the CO cores (tiny green circles). Colorbar values can be converted from counts to calibrated AB magnitudes with the equation: FUV$_{AB} = -2.5$log$_{10}($FUV$_{counts}) + 18.82.$ \label{fig:fuvWreg}}
\end{figure*}

\begin{figure*}[htb!]
\epsscale{1.1}\plotone{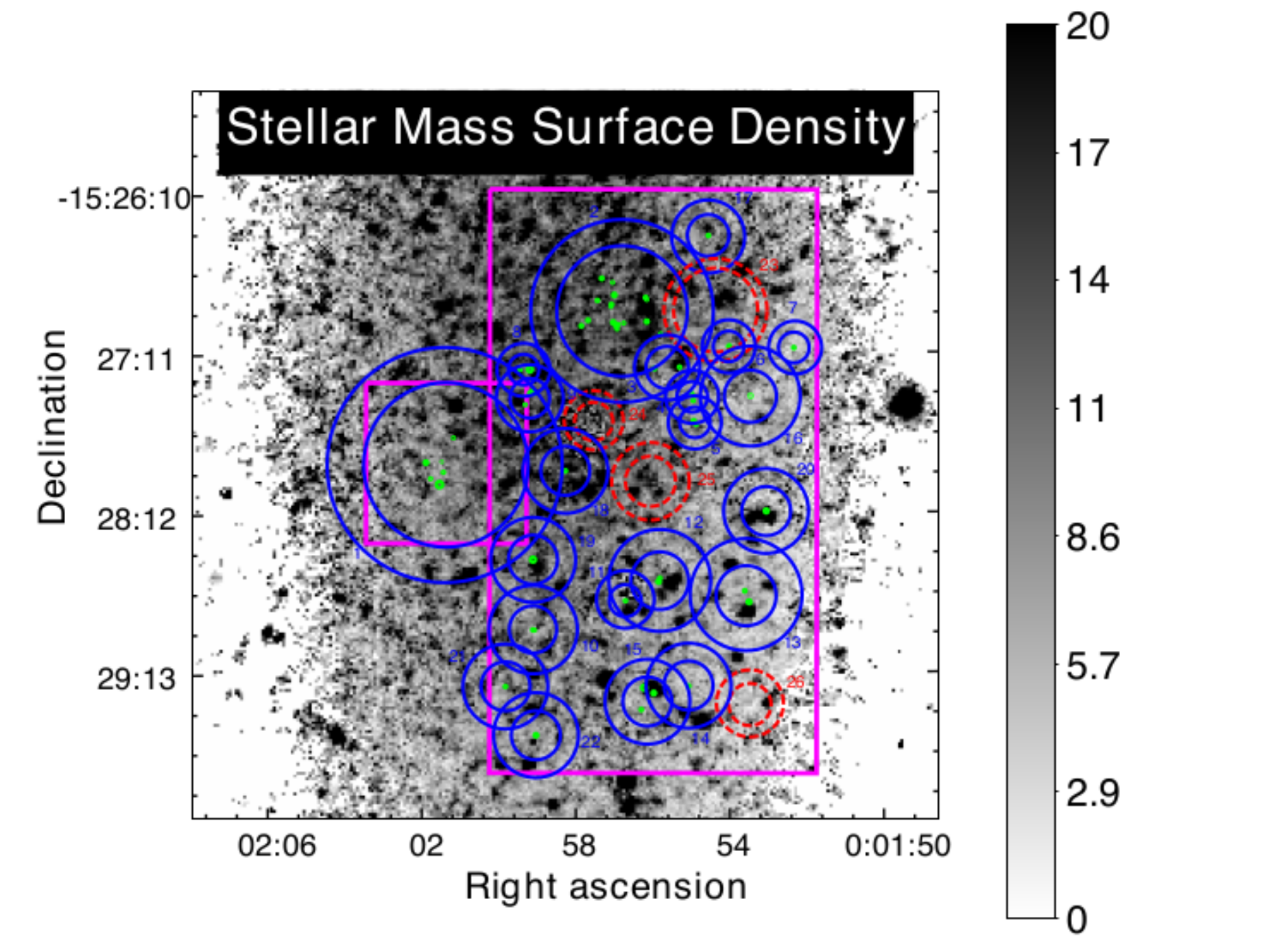}
\caption{As in Figure \ref{fig:fuvWreg}, but for stellar mass surface density. The colorbar values are in units of $M_\odot$ pc$^{-2}$. \label{fig:massWreg}}
\end{figure*}

\begin{figure*}[htb!]
\epsscale{1.1}\plotone{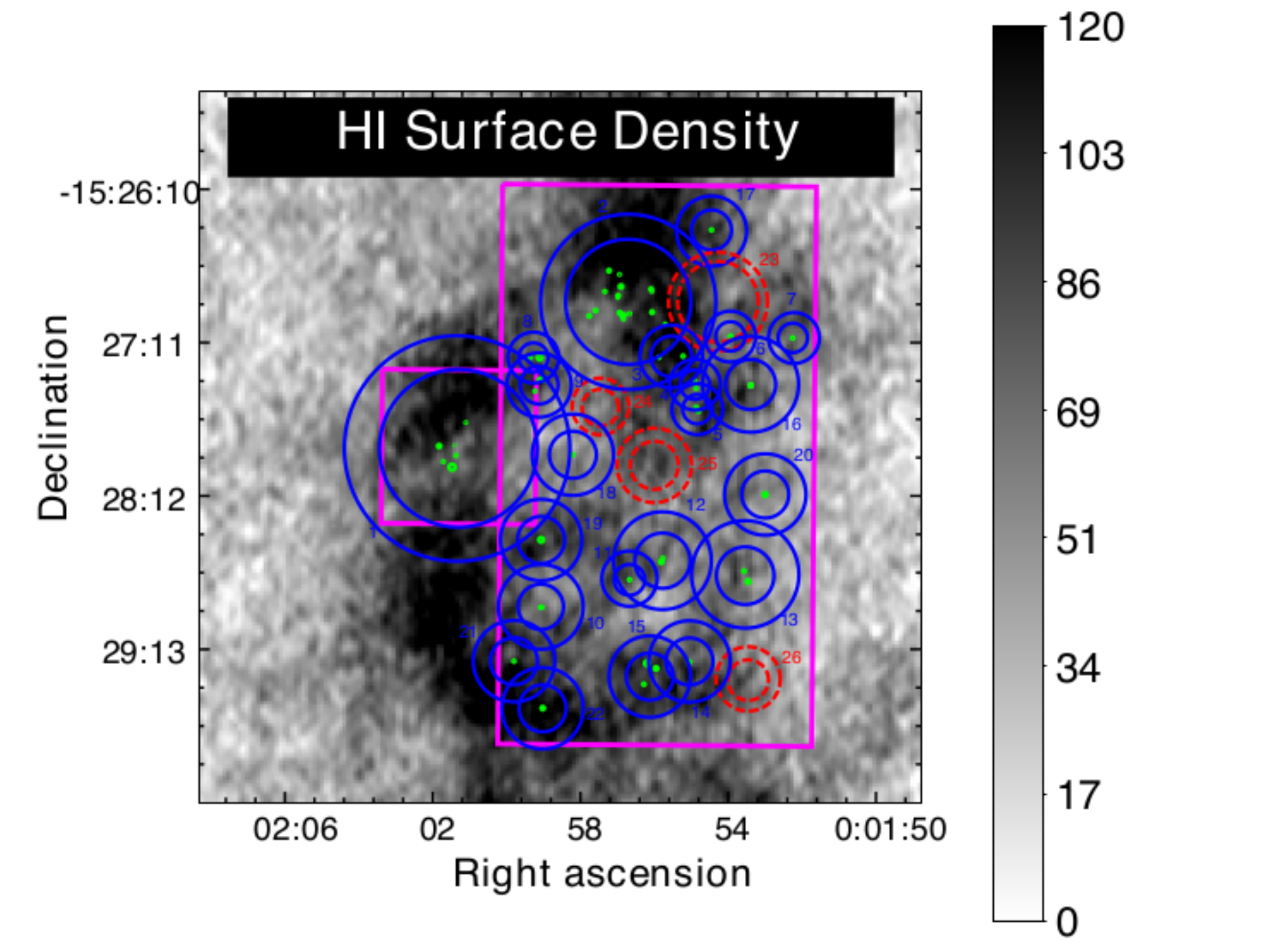}
\caption{As in Figure \ref{fig:fuvWreg}, but for \HI, with a beam size of $7\farcs6 \times 5\farcs1$. The colorbar values can be multiplied by 0.231 to convert from J bm$^{-1}$ m s$^{-1}$ to $M_\odot$ pc$^{-2}$ (see \S \ref{environ}).\label{fig:hiWreg}}
\end{figure*}

\begin{figure*}[htb!]
\epsscale{1.1}\plotone{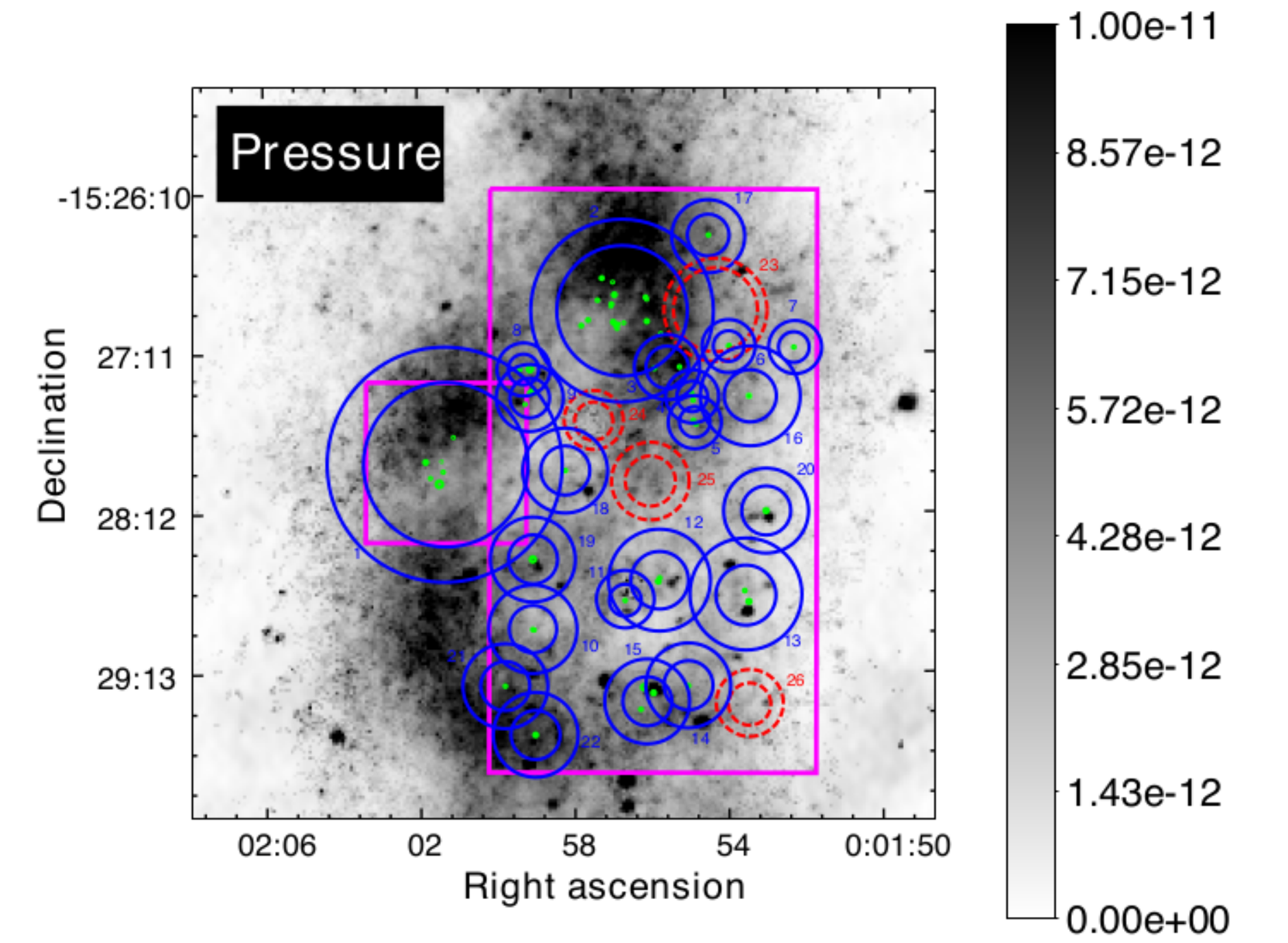}
\caption{As in Figure \ref{fig:fuvWreg}, but for pressure with a beam size of $7\farcs6 \times 5\farcs1$. The pressure is dominated by the gas, as is evident from a comparison with Figure \ref{fig:hiWreg}. The colorbar values are in units of g s$^{-2}$ cm$^{-1}$. \label{fig:presWreg}}
\end{figure*}

We attempted to cluster the CO cores using the clustering algorithm \textsc{DBSCAN} (Density-Based Spatial Clustering of Applications with Noise) \citep{DBLP:conf/kdd/EsterKSX96}. We found that we could not reproduce the known
clustering in region 1 with this algorithm. Depending on the parameters chosen, the algorithm would either include most CO cores across the galaxy in the same cluster or leave each core as an outlier.

Instead, we grouped the cores into 22 regions, several of which only include one CO core. Four additional regions (regions 23, 24, 25, and 26) were selected as regions including strong FUV emission without any CO cores. These regions were used as comparisons to the regions containing CO cores. The regions, along with the CO cores and ALMA FOV from both \citet{Rubio:2015} and Rubio et al.\ (2022, in preparation), are overlaid and labeled on the FUV, stellar mass surface density, \HI\ surface density, and pressure maps in Figures \ref{fig:fuvWreg}, \ref{fig:massWreg}, \ref{fig:hiWreg}, and \ref{fig:presWreg}, respectively. We list the region IDs, coordinates, and sizes
in Table \ref{tab:datatable1}.

\subsection{Environments}\label{environ}
Each region consists of a circular inner region representing the CO cores and the star-forming region in which they currently sit, along with an outer annulus which represents the projected environment in which the star-forming region formed. We will refer to the star forming regions as the \emph{inner regions} and the surrounding environments as the \emph{annuli}. For region 1, we chose the inner region to be about the size of the PDR. The outer annuli widths used in measuring the environmental characteristics ranged from 3.7 to 14.2 arcseconds depending on the location of the CO cores in the galaxy and their surrounding to minimize contamination from other regions. The annuli are also overlaid on the FUV, stellar mass surface density, \HI\ surface density, and presssure maps in Figures \ref{fig:fuvWreg}, \ref{fig:massWreg}, \ref{fig:hiWreg}, and \ref{fig:presWreg}, respectively and can be found in Table \ref{tab:datatable1}.

We used the Image Reduction and Analysis Facility (\textsc{IRAF}) \citep{Tody_1986} routine \textsc{Apphot} to measure the fluxes in the \HI\ surface density, stellar mass surface density, pressure map, $U$, $B$, $V$, FUV, and NUV images. We found the average pixel value for each inner region and the modal pixel value each outer annulus.
The modal pixel value was chosen for the annulus to minimize contamination from the environmental annulus overlapping other regions. We converted the pixel values for the \textit{UBV} images to Johnson magnitudes for each region and the FUV and NUV image values for each region to AB magnitudes. We also converted the \HI\ surface density values from Jy$\cdot$m beam$^{-1}$ s$^{-1}$ to $M_\odot$ pc$^{-2}$ using the equation
\begin{equation}
    M_{HI}=235.6\ D^2 \sum_{i} S_i\Delta V
\end{equation}
where $D$ is the distance to the object in Mpc, $S_i$ is the flux in Jy, $\Delta V$ is the velocity resolution in m s$^{-1}$, and $M_{HI}$ is in units of $M_\odot$ (Brinks, private communication). The Robust-weighted \HI\ moment 0 map ($S_i\Delta V$) is in units of Jy$\cdot$m beam$^{-1}$ s$^{-1}$ per pixel, so we divide this by the number of pixels per beam

\begin{equation}
    M_{HI}=235.6\ D^2 \sum_{i} S_i\Delta V\ \times \frac{pixel\ size^2}{1.13\ \Delta\alpha\Delta\delta}
    = 208.5\ D^2 \sum_{i} S_i\Delta V\ \times \frac{pixel\ size^2}{\Delta\alpha\Delta\delta}
\end{equation}

where the distance to WLM is 1 Mpc, the pixel size is 1\farcs5 and $\Delta\alpha$ and $\Delta\delta$ are the beam semi-major and semi-minor axis, which are 7\farcs6 and 5\farcs1 respectively. This makes the M$_{HI}$ in one pixel
\begin{equation}
    M_{HI}=12.21\ S[Jy/beam\cdot m/s/pixel]\ M_\odot.
\end{equation}
One pixel is 52.93 pc$^2$, so the \HI\ surface density becomes
\begin{equation}
    \Sigma_{HI}=0.231\ S[Jy/beam\cdot m/s/pixel]\ \ M_\odot/pc^2.
\end{equation}
The stellar mass density and pressure map images were already in units of $M_\odot$ pc$^{-2}$ and g s$^{-2}$ cm$^{-1}$ respectively. We list the center coordinates, radius of star forming region, and width of annulus of each region, along with the background subtracted and extinction corrected colors described in \S \ref{agemethod} for each inner region can be found in Table \ref{tab:datatable1}, while the inner region \HI\ surface densities and stellar mass surface densities can be found in Table \ref{tab:datatable2}. The extinction corrected colors, \HI\ surface densities, stellar mass surface densities, and pressures for each region's corresponding annulus can be found in Table \ref{tab:datatable3}. The quantities are averaged over regions or annuli that are larger than the 6\arcsec\ resolution of the \HI\ and pressure maps except for a few annuli widths, as can be seen in Table \ref{tab:datatable1}.

\begin{deluxetable*}{c|ccccrrrrr}
\centerwidetable
\tabletypesize{\scriptsize}
\tablecaption{Region centers \& background subtracted photometry of star forming regions \label{tab:datatable1}}
\tablenum{1}

\tablehead{\colhead{} & \colhead{} & \colhead{} & \colhead{} & \colhead{} & \colhead{} & \colhead{} & \colhead{} & \colhead{} & \colhead{} \\[-10pt]
\colhead{} & \colhead{} & \colhead{} & \colhead{Inner} & \colhead{Annulus} & \colhead{} & \colhead{} & \colhead{} & \colhead{} & \colhead{} \\[-8pt]
\colhead{} & \colhead{RA} & \colhead{Dec} & \colhead{Radius} & \colhead{Width} & \colhead{\textsl{FUV}} & \colhead{\textsl{FUV\,$-$\,NUV}} & \colhead{$V$} & \colhead{$U-B$} & \colhead{$B-V$}\\[-8pt]
\colhead{Region} & \colhead{\textsl{(J2000)}} & \colhead{\textsl{(J2000)}} & \colhead{(\arcsec)} & \colhead{(\arcsec)} & \colhead{\textsl{(mag)}} & \colhead{\textsl{(mag)}} & \colhead{\textsl{(mag)}} & \colhead{\textsl{(mag)}} & \colhead{\textsl{(mag)}} \\[-6pt]
\colhead{(1)} & \colhead{(2)} & \colhead{(3)} & \colhead{(4)} & \colhead{(5)} & \colhead{(6)} & \colhead{(7)} & \colhead{(8)}& \colhead{(9)} & \colhead{(10)}}

\startdata
1 & 0:02:01.7 & --15:27:53.0 & 23.4 & 10.2 & 22.61 $\pm$ 0.01 & --0.32 $\pm$ 0.01 & 23.88 $\pm$ 0.06 & --0.32 $\pm$ 0.14 & --0.17 $\pm$ 0.08 \\
2 & 0:01:57.0 & --15:26:55.5 & 25.0 & 10.0 & 23.27 $\pm$ 0.01 & --0.01 $\pm$ 0.01 & 23.66 $\pm$ 0.04 & --0.51 $\pm$ 0.09 & --0.06 $\pm$ 0.06 \\
3 & 0:01:55.8 & --15:27:16.2 & ~~8.2 & ~~4.4 & 23.35 $\pm$ 0.01 & 0.23 $\pm$ 0.01 & 23.60 $\pm$ 0.04 & --0.45 $\pm$ 0.09 & --0.11 $\pm$ 0.05 \\
4 & 0:01:55.1 & --15:27:27.5 & ~~5.9 & ~~4.3 & 24.54 $\pm$ 0.01 & 0.10 $\pm$ 0.01 & 25.41 $\pm$ 0.10 & --0.78 $\pm$ 0.29 & --0.41 $\pm$ 0.12 \\
5 & 0:01:55.1 & --15:27:37.3 & ~~5.9 & ~~4.3 & 24.73 $\pm$ 0.01 & 0.10 $\pm$ 0.01 & 24.72 $\pm$ 0.07 & --0.20 $\pm$ 0.20 & 0.16 $\pm$ 0.10 \\
6 & 0:01:54.2 & --15:27:08.3 & ~~5.9 & ~~4.3 & 22.70 $\pm$ 0.01 & --0.08 $\pm$ 0.01 & 23.84 $\pm$ 0.08 & 0.73 $\pm$ 0.29 & --0.30 $\pm$ 0.11 \\
7 & 0:01:52.5 & --15:27:08.9 & ~~5.9 & ~~4.3 & 23.88 $\pm$ 0.01 & --0.56 $\pm$ 0.01 & 26.77 $\pm$ 0.18 & --0.62 $\pm$ 0.23 & --1.11 $\pm$ 0.20 \\
8 & 0:01:59.6 & --15:27:16.7 & ~~5.9 & ~~4.3 & 17.04 $\pm$ 0.01 & --0.22 $\pm$ 0.01 & 21.69 $\pm$ 0.08 & --0.09 $\pm$ 0.33 & --0.01 $\pm$ 0.13 \\
9 & 0:01:59.5 & --15:27:27.7 & ~~7.6 & ~~5.2 & 22.73 $\pm$ 0.01 & --0.51 $\pm$ 0.01 & 23.97 $\pm$ 0.06 & --0.46 $\pm$ 0.14 & --0.18 $\pm$ 0.08 \\
10 & 0:01:59.4 & --15:28:56.2 & ~~9.1 & ~~7.8 & 23.62 $\pm$ 0.01 & --0.27 $\pm$ 0.01 & 24.70 $\pm$ 0.07 & --0.55 $\pm$ 0.14 & --0.18 $\pm$ 0.09 \\
11 & 0:01:56.0 & --15:28:45.4 & ~~6.1 & ~~4.9 & 25.34 $\pm$ 0.01 & 0.56 $\pm$ 0.01 & 24.69 $\pm$ 0.07 & --0.80 $\pm$ 0.14 & --0.04 $\pm$ 0.09 \\
12 & 0:01:56.1 & --15:28:37.8 & 11.0 & ~~9.2 & 22.17 $\pm$ 0.01 & --0.24 $\pm$ 0.01 & 23.20 $\pm$ 0.04 & --0.55 $\pm$ 0.08 & --0.08 $\pm$ 0.05 \\
13 & 0:01:53.8 & --15:28:43.8 & 11.5 & 11.3 & 24.00 $\pm$ 0.01 & 0.19 $\pm$ 0.01 & 23.48 $\pm$ 0.04 & --0.07 $\pm$ 0.11 & 0.05 $\pm$ 0.05 \\
14 & 0:01:55.3 & --15:29:18.1 & ~~9.4 & ~~6.8 & 25.22 $\pm$ 0.01 & --0.68 $\pm$ 0.01 & 26.15 $\pm$ 0.14 & --1.15 $\pm$ 0.22 & --0.31 $\pm$ 0.18 \\
15 & 0:01:56.4 & --15:29:24.1 & ~~9.4 & ~~6.8 & 22.68 $\pm$ 0.01 & --0.05 $\pm$ 0.01 & 23.29 $\pm$ 0.04 & --0.53 $\pm$ 0.10 & 0.19 $\pm$ 0.05 \\
16 & 0:01:53.6 & --15:27:27.4 & 10.0 & 10.0 & 23.13 $\pm$ 0.01 & --0.09 $\pm$ 0.01 & 23.19 $\pm$ 0.04 & --0.49 $\pm$ 0.09 & 0.16 $\pm$ 0.05 \\
17 & 0:01:54.7 & --15:26:25.8 & ~~8.0 & ~~6.0 & 24.60 $\pm$ 0.01 & --0.01 $\pm$ 0.01 & 24.75 $\pm$ 0.07 & --0.03 $\pm$ 0.18 & --0.12 $\pm$ 0.09 \\
18 & 0:01:58.5 & --15:27:55.5 & ~~9.4 & ~~6.8 & 19.77 $\pm$ 0.01 & \nodata \tablenotemark{\scriptsize{1}} & 22.80 $\pm$ 0.11 & 0.53 $\pm$ 0.36 & --0.31 $\pm$ 0.18 \\
19 & 0:01:59.4 & --15:28:29.5 & ~~9.4 & ~~6.8 & 23.20 $\pm$ 0.01 & --0.18 $\pm$ 0.01 & 23.99 $\pm$ 0.06 & --0.20 $\pm$ 0.16 & 0.05 $\pm$ 0.08 \\
20 & 0:01:53.2 & --15:28:11.4 & ~~9.4 & ~~6.8 & 17.90 $\pm$ 0.01 & --0.47 $\pm$ 0.01 & 21.81 $\pm$ 0.05 & --0.23 $\pm$ 0.19 & 0.07 $\pm$ 0.08 \\
21 & 0:02:00.2 & --15:29:17.9 & ~~9.4 & ~~6.8 & 24.66 $\pm$ 0.01 & --0.35 $\pm$ 0.01 & 25.28 $\pm$ 0.09 & --0.07 $\pm$ 0.21 & --0.29 $\pm$ 0.12 \\
22 & 0:01:59.4 & --15:29:36.7 & ~~9.4 & ~~6.8 & \nodata \tablenotemark{\scriptsize{2}} & \nodata  \tablenotemark{\scriptsize{2}}& 23.98 $\pm$ 0.08 & 1.06 $\pm$ 0.37 & --0.28 $\pm$ 0.12 \\
23 & 0:01:54.5 & --15:26:54.4 & 16.0 & 14.2 & 23.08 $\pm$ 0.01 & --0.27 $\pm$ 0.01 & 24.11 $\pm$ 0.06 & --0.71 $\pm$ 0.13 & --0.12 $\pm$ 0.08 \\
24 & 0:01:57.8 & --15:27:36.7 & ~~7.4 & ~~4.0 & 24.90 $\pm$ 0.01 & --0.07 $\pm$ 0.01 & 28.97 $\pm$ 0.50 & --4.94 $\pm$ 0.25 & 1.45 $\pm$ 0.55 \\
25 & 0:01:56.3 & --15:27:59.8 & ~~9.6 & ~~5.3 & 22.48 $\pm$ 0.01 & --0.08 $\pm$ 0.01 & 22.97 $\pm$ 0.04 & --0.18 $\pm$ 0.14 & --0.08 $\pm$ 0.06 \\
26 & 0:01:53.7 & --15:29:25.4 & ~~8.1 & ~~3.7 & 20.62 $\pm$ 0.01 & --0.40 $\pm$ 0.01 & 23.58 $\pm$ 0.06 & --0.06 $\pm$ 0.23 & --0.23 $\pm$ 0.11 \\
\enddata

\tablenotetext{1}{NUV background counts were higher than measured inside the region.}

\tablenotetext{2}{\textsl{FUV} background counts were higher than measured inside the region.}

\tablecomments{Magnitudes and colors are background subtracted and corrected for Galactic foreground and internal extinction.}

\end{deluxetable*}

\begin{deluxetable*}{c|ccccccccc}
\centerwidetable
\tabletypesize{\scriptsize}
\tablecaption{Characteristics of star forming regions \label{tab:datatable2}}
\tablenum{2}

\tablehead{\colhead{} & \colhead{} & \colhead{} & \colhead{} & \colhead{} & \colhead{} & \colhead{} & \colhead{} & \colhead{} \\[-10pt]
\colhead{} & \colhead{} & \colhead{} & \colhead{} & \colhead{} & \colhead{} & \colhead{} & \colhead{} & \colhead{Log Age} \\[-8pt]
\colhead{} & \colhead{$\Sigma_{*\tiny{\textnormal{,SF}}}$} & \colhead{$\Sigma_{\tiny{\textnormal{H}\,\textnormal{\sc{i},SF}}}$} & \colhead{$\Sigma_{\tiny{\textnormal{H}\,\textnormal{\sc{i},SF}}} - \Sigma_{\tiny{\textnormal{H}\,\textnormal{\sc{i}}\,\textnormal{,rad}}}$ \tablenotemark{\scriptsize{1}}} & \colhead{$M_{\tiny{\textrm{CO, }vir}}$ \tablenotemark{\scriptsize{2}}} & \colhead{$\Sigma_{\tiny{\textrm{CO}}}$ \tablenotemark{\scriptsize{3}}} & \colhead{$M_{\tiny{Dark\textrm{ H}_2}}$/$M_{\tiny{MC}}$ \tablenotemark{\scriptsize{4}}} & \colhead{$E(B\!-\!V)$} & \colhead{(Chabrier IMF)} \\ [-8pt]
\colhead{Region} & \colhead{\textsl{($M_\odot \textrm{ pc}^{-2}$)}} & \colhead{\textsl{($M_\odot \textrm{ pc}^{-2}$)}} & \colhead{\textsl{($M_\odot \textrm{ pc}^{-2}$)}} & \colhead{($M_\odot$)} & \colhead{\textsl{($M_\odot \textrm{ pc}^{-2}$)}} & \colhead{(\%)} & \colhead{(Chabrier IMF)} & \colhead{\textsl{(years)}} \\[-6pt]
\colhead{(1)} & \colhead{(2)} & \colhead{(3)} & \colhead{(4)} & \colhead{(5)} & \colhead{(6)} & \colhead{(7)} & \colhead{(8)} & \colhead{(9)}} 

\startdata
1 & $0.06^{+0.74}_{-0.00}$ & 21.33$\pm$0.01 & 15.30$\pm$0.01 & 21700$\pm$6440 & $106.4^{+421.6}_{-67.6}$ & $82.1^{+17.2}_{-0.0}$ & 0.21 & $6.62^{+1.19}_{-0.02}$ \\
2 & $0.91^{+ 0.03}_{-0.39}$ & 24.05$\pm$0.01 & 17.96$\pm$0.01 & 33900$\pm$5260 & $106.7^{+682.5}_{-80.0}$ & $98.4^{+0.8}_{-0.0}$ & 0.06 & $7.74^{+0.02}_{-0.42}$ \\
3 & $1.18^{+0.24}_{-0.19}$ & 21.91$\pm$0.01 & 16.13$\pm$0.01& 4300$\pm$2550 & $101.4^{+54.1}_{-54.1}$ & $98.5^{+0.9}_{-0.0}$ & 0.06 & $7.91^{+0.15}_{-0.15}$ \\
4 & $0.56^{+ 0.05}_{-0.21}$ & 20.11$\pm$0.01 & 14.54$\pm$0.01 & 1790$\pm$888 & $74.3^{+50.8}_{-50.8}$ & $97.5^{+1.4}_{-0.0}$ & 0.06 & $8.56^{+0.05}_{-0.30}$ \\
5 & $0.55^{+ 0.16}_{-0.21}$ & 19.63$\pm$0.01 & 14.61$\pm$0.01 & 4160$\pm$2000 & $145.1^{+105.8}_{-106.0}$ & $94.2^{+3.6}_{-0.0}$ & 0.08 & $8.16^{+0.20}_{-0.40}$ \\
6 & $1.46^{+ 0.39}_{-0.29}$ & 16.15$\pm$0.01 & 11.63$\pm$0.01 & 2140$\pm$1130 & $61.5^{+0}_{-0}$ & $98.9^{+0.7}_{-0.0}$ & 0.42 & $8.26^{+0.20}_{-0.15}$ \\
7 & $0.02^{+0.00}_{-0.01}$ & 16.90$\pm$0.01 & 13.80$\pm$0.01 & 571$\pm$1050 & $32.70^{+0}_{-0}$ & $71.4^{+14.3}_{-17.1}$ & 0.06 & $6.40^{+0.10}_{-0.00}$ \\
8 & $2.05^{+2.60}_{-0.00}$ & 18.61$\pm$0.01 & 15.26$\pm$0.01 & 2800$\pm$1960 & $121.1^{+61.1}_{-61.1}$ & $98.9^{+0.8}_{-0.0}$ & 1.10 & $7.10^{+0.53}_{-0.08}$ \\
9 & $0.06^{+ 0.00}_{-0.01}$ & 20.45$\pm$0.01 & 16.73$\pm$0.01 & 16800$\pm$2170 & $138.4^{+56.1}_{-77.6}$ & $0.0^{+38.7}_{-0.0}$ & 0.20 & $6.50^{+0.02}_{-0.00}$ \\
10 & $0.03^{+ 0.02}_{-0.00}$ & 15.48$\pm$0.01 & 11.32$\pm$0.01 & 3050$\pm$1900 & $100.3^{+34.7}_{-34.7}$ & $67.1^{+23.7}_{-0.0}$ & 0.06 & $6.64^{+0.02}_{-0.12}$ \\
11 & $0.21^{+0.46}_{-0.02}$ & 17.44$\pm$0.01 & 12.70$\pm$0.01 & 935$\pm$586 & $114.6^{+0}_{-0}$ & $96.8^{+0.2}_{-0.0}$ & 0.06 & $7.34^{+0.92}_{-0.16}$ \\
12 & $0.11^{+0.01}_{-0.00}$ & 15.62$\pm$0.01 & 10.70$\pm$0.01 & 6010$\pm$1820 & $85.4^{+31.5}_{-31.5}$ & $87.7^{+6.6}_{-0.0}$ & 0.08 & $6.62^{+0.00}_{-0.02}$ \\
13 & $2.05^{+ 0.33}_{-0.18}$ & 14.92$\pm$0.01 & 11.63$\pm$0.01 & 3960$\pm$2330 & $146.9^{+3.1}_{-6.1}$ & $99.6^{+0.2}_{-0.0}$ & 0.06 & $8.26^{+0.15}_{-0.05}$ \\
14 & $0.03^{+0.00}_{-0.01}$ & 19.05$\pm$0.01 & 15.58$\pm$0.01 & 701$\pm$565 & $33.3^{+0}_{-0}$ & $92.2^{+3.9}_{-0.0}$ & 0.08 & $6.40^{+0.06}_{-0.00}$ \\
15 & $1.03^{+ 0.00}_{-0.01}$ & 22.03$\pm$0.01 & 18.16$\pm$0.01 & 10100$\pm$5020 & $182.5^{+232.9}_{-97.6}$ & $97.0^{+1.6}_{-0.0}$ & 0.11 & $7.59^{+0.06}_{-0.15}$ \\
16 & $1.17^{+ 0.08}_{-0.25}$ & 17.84$\pm$0.01 & 13.09$\pm$0.01 & 543$\pm$1130 & $34.6^{+0}_{-0}$ & $99.9^{+0.1}_{-0.0}$ & 0.08 & $7.63^{+0.05}_{-0.17}$ \\
17 & $0.46^{+0.12}_{-0.05}$ & 21.02$\pm$0.01 & 16.13$\pm$0.01 & 1760$\pm$2780 & $64.4^{+0}_{-0}$ & $98.4^{+1.0}_{-0.0}$ & 0.06 & $8.01^{+0.20}_{-0.10}$ \\
18 & $0.60^{+0.19}_{-0.05}$ & 12.71$\pm$0.01 & ~~8.51$\pm$0.01 & 469$\pm$1090 & $30.3^{+0}_{-0}$ & $99.8^{+0.3}_{-0.0}$ & 1.03 & $6.40^{+0.72}_{-0.00}$ \\
19 & $0.73^{+ 0.14}_{-0.05}$ & 20.98$\pm$0.01 & 17.09$\pm$0.01 & 1020$\pm$1680 & $102.4^{+0}_{-0}$ & $99.6^{+0.2}_{-0.0}$ & 0.16 & $8.01^{+0.20}_{-0.10}$ \\
20 & $0.47^{+ 0.66}_{-0.00}$ & 12.78$\pm$0.01 & ~~9.69$\pm$0.01 & 518$\pm$1010 & $146.5^{+0}_{-0}$ & $99.7^{+0.3}_{-0.0}$ & 0.81 & $6.50^{+0.48}_{-0.02}$ \\
21 & $0.25^{+ 0.01}_{-0.23}$ & 24.15$\pm$0.01 & 19.58$\pm$0.01 & 447$\pm$634 & $43.8^{+0}_{-0}$ & $99.5^{+0.3}_{-3.7}$ & 0.06 & $7.91^{+0.05}_{-1.29}$ \\
22 & $2.68^{+0.00}_{-0.00}$ & 26.71$\pm$0.01 & 22.06$\pm$0.01 & ~~665$\pm$1000 & $94^{+0}_{-0}$ & $99.9^{+0.0}_{-0.0}$ & 0.43 & $8.71^{+0.00}_{-0.00}$ \\
23 & $0.09^{+ 0.02}_{-0.00}$ & 18.73$\pm$0.01 & 13.02$\pm$0.01 & 0 & 0 & 100 & 0.17 & $6.76^{+0.08}_{-0.04}$ \\
24 & $0.002^{+0.000}_{-0.000}$ & 15.22$\pm$0.01 & 10.54$\pm$0.01 & 0 & 0 & 100 & 0.06 & $6.40^{+0.00}_{-0.00}$ \\
25 & $2.32^{+ 1.43}_{-0.00}$ & ~~7.52$\pm$0.01 & ~~1.67$\pm$0.01 & 0 & 0 & 100 & 0.30 & $8.01^{+0.40}_{-0.15}$ \\
26 & $0.08^{+ 1.42}_{-0.00}$ & 15.64$\pm$0.01 & 12.86$\pm$0.01 & 0 & 0 & 100 & 0.51 & $6.62^{+1.49}_{-1.19}$ \\
\enddata
\tablenotetext{1}{Average star forming region $\Sigma_{\tiny{\textnormal{H}\,\textnormal{\sc{i}}\,\textnormal{,SF}}}$ above radial average $\Sigma_{\tiny{\textnormal{H}\,\textnormal{\sc{i}}\,\textnormal{,rad}}}$. Uncertainties are carried over from the $\Sigma_{\tiny{\textnormal{H}\,\textnormal{\sc{i}}\,\textnormal{,SF}}}$ of the region.}
\tablenotetext{2}{Sum of CO core virial masses given by \citet{Rubio:2015} and Rubio et al.\ (2022, in preparation)}
\tablenotetext{3}{\citet{Rubio:2015} and Rubio et al.\ (2022, in preparation) calculated the $\Sigma_{\tiny{\textrm{CO}}}$ of the individual CO cores from their $M_{\tiny{\textrm{CO},vir}}$. For each ensemble of individual $\Sigma_{\tiny{\textrm{CO}}}$ in a given region, we adopt the median $\Sigma_{\tiny{\textrm{CO}}}$ value. The $\Sigma_{\tiny{\textrm{CO}}}$ $\pm$ uncertainties represent the range of $\Sigma_{\tiny{\textrm{CO}}}$ for each region containing multiple CO cores.}
\tablenotetext{4}{Percentage of original total molecular cloud mass in dark H$_2$.}

\end{deluxetable*}

\begin{deluxetable*}{c|ccccccccc}
\centerwidetable

\tabletypesize{\scriptsize}
\tablecaption{Colors and characteristics of environments \label{tab:datatable3}}
\tablenum{3}

\tablehead{\colhead{} & \colhead{} & \colhead{} & \colhead{} & \colhead{} & \colhead{} & \colhead{} & \colhead{} & \colhead{} & \colhead{} \\[-10pt]
\colhead{} & \colhead{} & \colhead{} & \colhead{} & \colhead{} & \colhead{} & \colhead{} & \colhead{Pressure} & \colhead{} & \colhead{Log Age} \\[-8pt]
\colhead{} & \colhead{\textsl{FUV}} & \colhead{\textsl{FUV\,$-$\,NUV}} & \colhead{$V$} & \colhead{$U-B$} & \colhead{$B-V$} & \colhead{$\Sigma_{*\tiny{\textnormal{,env}}}$} & \colhead{($\times 10^{-12}$)} & \colhead{$\Sigma_{\tiny{\textnormal{H}\,\textnormal{\sc{i}}\,\textnormal{,env}}}$} & \colhead{(Chabrier IMF)} \\[-8pt]
\colhead{Region} & \colhead{\textsl{(mag)}} & \colhead{\textsl{(mag)}} & \colhead{\textsl{(mag)}} & \colhead{\textsl{(mag)}} & \colhead{\textsl{(mag)}} & \colhead{\textsl{($M_\odot \textrm{ pc}^{-2}$)}} & \colhead{\textsl{(g s$^{-2}$ cm$^{-1}$)}} & \colhead{\textsl{($M_\odot \textrm{ pc}^{-2}$)}} & \colhead{\textsl{(years)}} \\[-6pt]
\colhead{(1)} & \colhead{(2)} & \colhead{(3)} & \colhead{(4)} & \colhead{(5)} & \colhead{(6)} & \colhead{(7)} & \colhead{(8)} & \colhead{(9)} & \colhead{(10)}} 

\startdata
1 & 24.87$\pm$0.01 & ~~0.58$\pm$0.01 & 22.86$\pm$0.03 & --0.04$\pm$0.10 & 0.36$\pm$0.04 & 10.30$\pm$0.28 & 3.86$\pm$4.29 & 19.50$\pm$0.01 & $8.81^{+0.00}_{-0.05}$ \\
2 & 24.43$\pm$0.01 & ~~0.26$\pm$0.01 & 22.65$\pm$0.03 & 0.09$\pm$0.09 & 0.37$\pm$0.04 & 12.02$\pm$0.29 & 5.66$\pm$4.67 & 21.22$\pm$0.01 & $8.61^{+0.05}_{-0.15}$ \\
3 & 23.69$\pm$0.01 & ~~0.01$\pm$0.01 & 22.52$\pm$0.03 & --0.08$\pm$0.08 & 0.28$\pm$0.04 & 12.38$\pm$0.28 & 5.68$\pm$4.02 & 18.27$\pm$0.01 & $8.31^{+0.05}_{-0.20}$ \\
4 & 23.84$\pm$0.01 & ~~0.03$\pm$0.01 & 22.87$\pm$0.03 & --0.17$\pm$0.09 & 0.32$\pm$0.04 & ~~9.10$\pm$0.25 & 5.33$\pm$4.80 & 21.81$\pm$0.01 & $8.16^{+0.05}_{-0.20}$ \\
5 & 24.09$\pm$0.01 & --0.02$\pm$0.01 & 23.10$\pm$0.03 & 0.04$\pm$0.11 & 0.31$\pm$0.05 & ~~8.79$\pm$0.27 & 4.78$\pm$4.17 & 18.92$\pm$0.01 & $8.36^{+0.05}_{-0.20}$ \\
6 & 24.16$\pm$0.01 & ~~0.18$\pm$0.01 & 23.00$\pm$0.03 & --0.002$\pm$0.10 & 0.33$\pm$0.05 & ~~8.25$\pm$0.24 & 3.59$\pm$3.67 & 16.66$\pm$0.01 & $8.46^{+0.00}_{-0.05}$ \\
7 & 24.74$\pm$0.01 & ~~0.29$\pm$0.01 & 23.31$\pm$0.04 & 0.06$\pm$0.12 & 0.30$\pm$0.05 & ~~5.89$\pm$0.19 & 2.81$\pm$3.42 & 15.56$\pm$0.01 & $8.61^{+0.05}_{-0.10}$ \\
8 & 23.85$\pm$0.01 & ~~0.15$\pm$0.01 & 22.46$\pm$0.02 & 0.08$\pm$0.08 & 0.25$\pm$0.04 & 13.83$\pm$0.31 & 5.56$\pm$4.19 & 19.05$\pm$0.01 & $8.46^{+0.00}_{-0.05}$ \\
9 & 23.94$\pm$0.01 & ~~0.31$\pm$0.01 & 22.45$\pm$0.02 & 0.02$\pm$0.08 & 0.35$\pm$0.04 & 13.65$\pm$0.31 & 5.07$\pm$4.13 & 18.78$\pm$0.01 & $8.61^{+0.05}_{-0.10}$ \\
10 & 24.44$\pm$0.01 & ~~0.09$\pm$0.01 & 22.87$\pm$0.03 & --0.02$\pm$0.10 & 0.36$\pm$0.04 & 10.11$\pm$0.27 & 4.38$\pm$3.94 & 17.90$\pm$0.01 & $8.41^{+0.05}_{-0.05}$ \\
11 & 24.53$\pm$0.01 & ~~0.04$\pm$0.01 & 22.87$\pm$0.03 & --0.03$\pm$0.10 & 0.50$\pm$0.05 & ~~8.68$\pm$0.23 & 3.65$\pm$4.08 & 18.56$\pm$0.01 & $8.41^{+0.05}_{-0.05}$ \\
12 & 24.18$\pm$0.01 & ~~0.18$\pm$0.01 & 22.88$\pm$0.03 & --0.09$\pm$0.09 & 0.31$\pm$0.04 & ~~7.67$\pm$0.21 & 2.48$\pm$3.01 & 13.71$\pm$0.01 & $8.41^{+0.05}_{-0.05}$ \\
13 & 23.12$\pm$0.01 & --0.08$\pm$0.01 & 23.05$\pm$0.03 & --0.36$\pm$0.08 & 0.18$\pm$0.05 & ~~5.04$\pm$0.15 & 1.93$\pm$2.75 & 12.50$\pm$0.01 & $7.72^{+0.09}_{-0.02}$ \\
14 & 23.77$\pm$0.01 & ~~0.14$\pm$0.01 & 23.01$\pm$0.03 & --0.14$\pm$0.09 & 0.22$\pm$0.05 & ~~6.98$\pm$0.20 & 3.39$\pm$4.16 & 18.89$\pm$0.01 & $8.31^{+0.10}_{-0.10}$ \\
15 & 24.91$\pm$0.01 & ~~0.14$\pm$0.01 & 23.01$\pm$0.03 & --0.21$\pm$0.09 & 0.28$\pm$0.05 & ~~7.64$\pm$0.22 & 4.97$\pm$4.58 & 20.80$\pm$0.01 & $8.31^{+0.10}_{-0.10}$ \\
16 & 24.48$\pm$0.01 & ~~0.19$\pm$0.01 & 23.25$\pm$0.04 & --0.15$\pm$0.11 & 0.36$\pm$0.05 & ~~6.40$\pm$0.20 & 2.86$\pm$3.21 & 14.61$\pm$0.01 & $8.41^{+0.05}_{-0.05}$ \\
17 & 25.51$\pm$0.01 & ~~0.11$\pm$0.01 & 23.13$\pm$0.03 & 0.18$\pm$0.12 & 0.49$\pm$0.05 & ~~9.24$\pm$0.28 & 4.97$\pm$4.39 & 19.97$\pm$0.01 & $8.46^{+0.00}_{-0.00}$ \\
18 & 24.46$\pm$0.01 & ~~0.39$\pm$0.01 & 22.60$\pm$0.03 & 0.19$\pm$0.09 & 0.41$\pm$0.04 & 13.33$\pm$0.32 & 2.96$\pm$2.84 & 12.90$\pm$0.01 & $8.71^{+0.05}_{-0.05}$ \\
19 & 23.96$\pm$0.01 & ~~0.19$\pm$0.01 & 22.71$\pm$0.03 & --0.06$\pm$0.09 & 0.32$\pm$0.04 & 10.54$\pm$0.26 & 4.48$\pm$4.00 & 18.17$\pm$0.01 & $8.41^{+0.05}_{-0.05}$ \\
20 & 25.33$\pm$0.01 & ~~0.37$\pm$0.01 & 23.63$\pm$0.04 & 0.11$\pm$0.14 & 0.38$\pm$0.06 & ~~6.18$\pm$0.24 & 1.71$\pm$2.55 & 11.58$\pm$0.01 & $8.66^{+0.10}_{-0.00}$ \\
21 & 24.83$\pm$0.01 & ~~0.37$\pm$0.01 & 22.99$\pm$0.03 & 0.004$\pm$0.10 & 0.41$\pm$0.05 & ~~8.66$\pm$0.25 & 7.54$\pm$5.09 & 23.11$\pm$0.01 & $8.66^{+0.05}_{-0.05}$ \\
22 & 23.89$\pm$0.01 & ~~0.09$\pm$0.01 & 22.92$\pm$0.03 & --0.17$\pm$0.09 & 0.28$\pm$0.04 & ~~7.92$\pm$0.22 & 7.25$\pm$5.33 & 24.21$\pm$0.01 & $8.31^{+0.05}_{-0.20}$ \\
23 & 24.64$\pm$0.01 & ~~0.36$\pm$0.01 & 22.98$\pm$0.03 & 0.22$\pm$0.11 & 0.39$\pm$0.05 & ~~8.74$\pm$0.30 & 3.40$\pm$4.36 & 19.81$\pm$0.01 & $8.71^{+0.00}_{-0.05}$ \\
24 & 20.79$\pm$0.01 & ~~0.12$\pm$0.01 & 22.57$\pm$0.03 & 0.10$\pm$0.09 & 0.36$\pm$0.04 & 11.89$\pm$0.21 & 4.35$\pm$3.60 & 16.37$\pm$0.01 & $8.46^{+0.00}_{-0.05}$ \\
25 & 21.02$\pm$0.01 & ~~0.29$\pm$0.01 & 22.89$\pm$0.03 & --0.07$\pm$0.09 & 0.31$\pm$0.04 & 10.03$\pm$0.27 & 3.77$\pm$3.71 & 16.84$\pm$0.01 & $8.51^{+0.10}_{-0.00}$ \\
26 & 21.83$\pm$0.01 & ~~0.45$\pm$0.01 & 21.57$\pm$0.03 & 0.10$\pm$0.09 & 0.36$\pm$0.04 & ~~4.68$\pm$0.12 & 2.79$\pm$3.36 & 15.26$\pm$0.01 & $8.76^{+0.00}_{-0.05}$ \\
\enddata

\tablecomments{Magnitudes and colors have been corrected for extinction.}
\end{deluxetable*}

\subsection{Age}\label{agemethod}
To calculate the age of each inner region,
we used the colors in the region and iterated on reddening to find the best fit with cluster evolutionary models.
First, to determine the region colors, we subtracted the background stellar disk from each region. To do so, we subtracted the mode of the pixel values of the outer annulus, measured using \textsc{Apphot}, from the average pixel value in the inner region. We chose to use the mode of the surroundings rather than the average in order minimize contamination from other star-forming regions (partially) sampled by the annulus.

\begin{figure*}[htb!]
\epsscale{1.05}\plotone{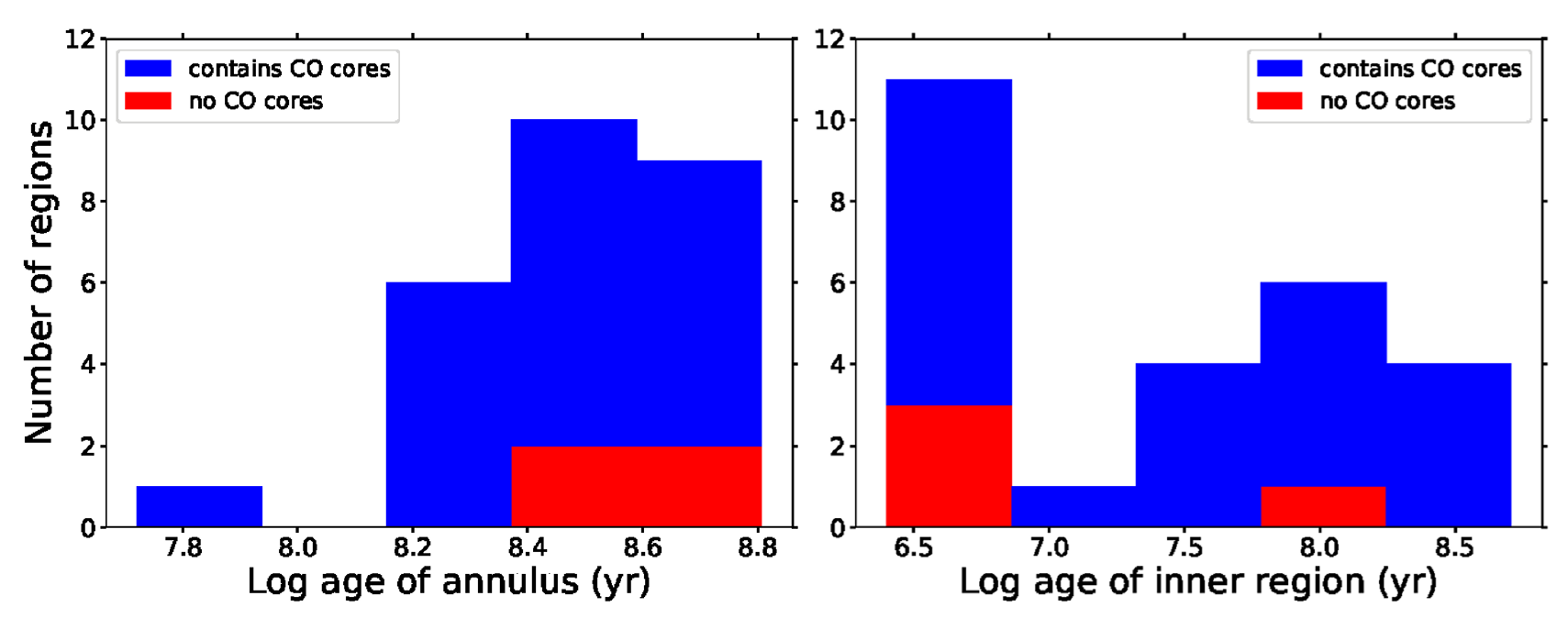}
\caption{Left: Histogram of ages of the environment where CO cores formed (blue) and environment surrounding the star forming regions without any CO cores (red). Right: Histogram of ages of star forming regions with CO cores (blue) and the star forming regions without any CO cores (red). Ages were computed using the Chabrier IMF.  \label{fig:ageHis}}
\end{figure*}

To find the extinction toward each region, we computed \textsl{FUV\,$-$\,NUV}, $U-B$, and $B-V$ colors using a series of $E(B\!-\!V)$ values ranging from 0.06--1.5 in steps of 0.01. The lower $E(B\!-\!V)$ limit of 0.06 was selected from adding the Milky Way foreground reddening and a minimal (0.05 mag) internal reddening for stars \citep{Schlafly_2011, Cardelli_1989}. We used an upper limit $E(B\!-\!V)$ of 1.5 in our model search as it is higher than the estimates of reddening necessary to form CO molecules in the Milky Way \citep{glover2012molecular,Lee_2018}.

We then compared the background-subtracted \textsl{FUV\,$-$\,NUV}, $U-B$, and $B-V$ colors to evolutionary stellar population synthesis models from GALEXEV \citep{Bruzual:2003aa}.  The single stellar population (SSP) models used were computed using the \citet{BERTELLI:1994aa} Padova evolutionary tracks with a metallicity of 0.004, as this was closest of the models to the WLM metallicity of $\sim$0.003. We compared models computed using both the \citet{Chabrier_2003} initial mass function (IMF) and the \citet{Salpeter_1955} IMF with the aforementioned evolutionary tracks and metallicity. We then compared the observed and modeled \textsl{FUV\,$-$\,NUV}, $U-B$, and $B-V$ colors to find the age that corresponded to the closest fit for each $E(B\!-\!V)$ value. For each region, we adopt the combination of age and $E(B\!-\!V)$ that minimizes the residuals between observed and modeled colors. Using this extinction, we then compared the \textsl{FUV\,$-$\,NUV}, $U-B$, and $B-V$ colors with their respective upper and lower uncertainty limits and the model colors, and selected these as the worst case scenario upper and lower age uncertainties of that region. Both IMFs produced similar results, so we report only $E(B\!-\!V)$ and ages calculated using the Chabrier IMF in Table \ref{tab:datatable2}. Figure \ref{fig:ageHis} shows a histogram of the ages of the inner regions.

\begin{figure*}[htb!]
\epsscale{1}\plotone{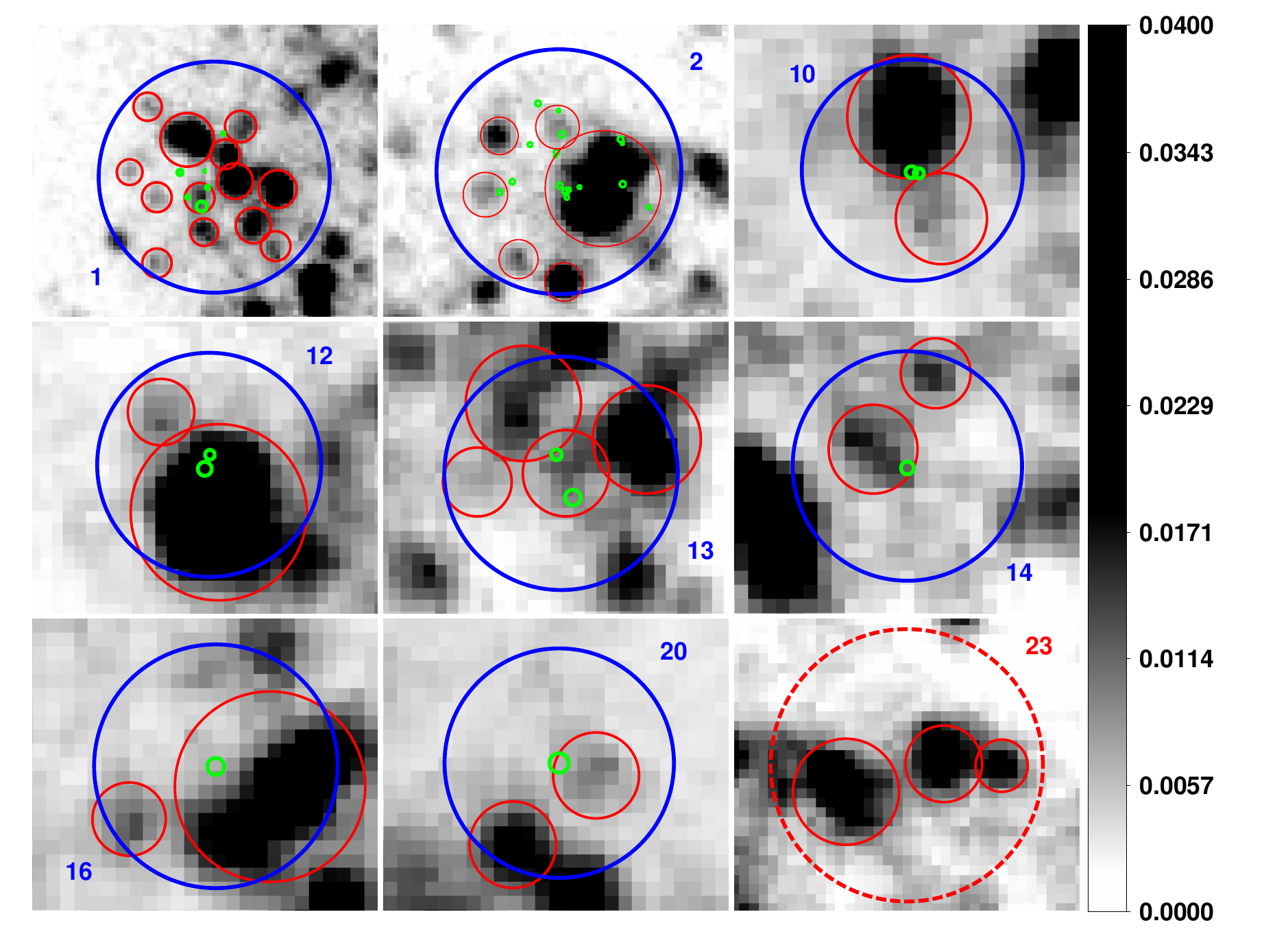}
\caption{FUV image of star forming regions (blue solid or red dashed circles), star forming knots (red circles), and CO cores (green circles) for which we compared the average of the ages of the individual knots to the calculated age of the entire region. We find that the individual knot ages are the same or similar to the age of the region comprising the knots. \label{fig:knots}}
\end{figure*}

\citet{Grasha_2018,Grasha_2019} find a strong correlation between the age of star clusters and distance from their GMCs, with the age distribution increasing as the cluster-GMC distance increases. For star forming regions that contained multiple FUV knots, we used the method described above to explore the ages of the individual knots compared to the age we computed for the entire region. We selected the photometry aperture size for each knot based on the FUV and $V$ images to encompass as much as was likely to be part of the same star forming knot. We used a larger aperture size around multiple knots that could not be individually resolved. For each star forming region, the individual knots have the same or similar ages as that of the entire region. For example, individual knots in region 1 range from 3.2 to 4.4 Myr, with 10 out of 13 regions having the same age as we calculated for the entire region -- 4.2 Myr. Similarly, the average $E(B\!-\!V)$ of the individual knots in region 1 is 0.23, while that of the entire region is 0.21. Figure \ref{fig:knots} shows the star forming regions with multiple knots and the photometry apertures we selected for the individual knots in each region. We note that the ages and reddening of the knots are sensitive to aperture size and background subtraction selection due to the crowding of knots within the regions.

\begin{figure*}[htb!]
\epsscale{0.8}\plotone{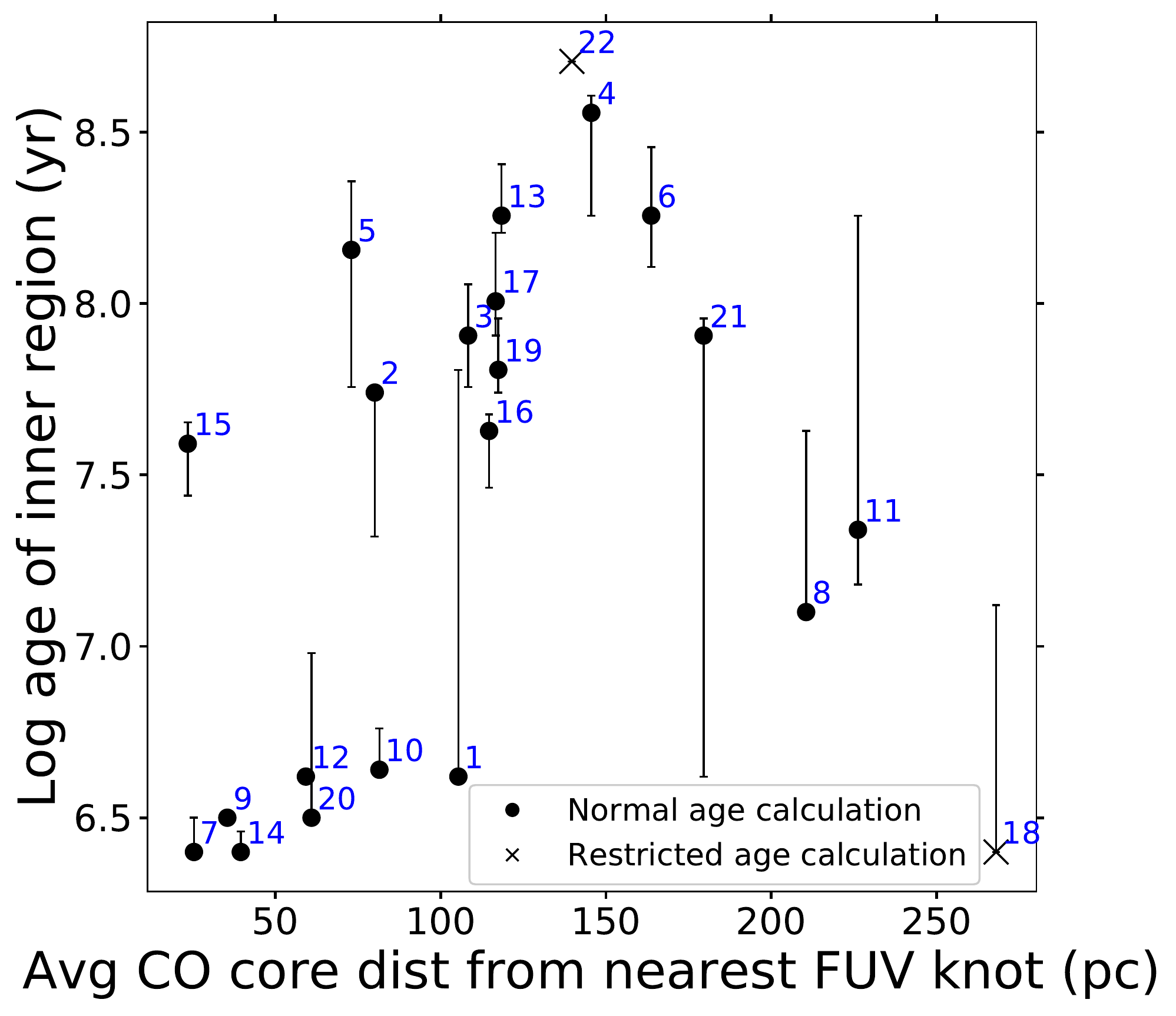}
\caption{The age of the inner region plotted against the average distance of the CO cores in that region to its nearest FUV knot in parsecs. Age error bars that appear one-sided are the result of the upper or lower color uncertainties finding the same age in the model. Regions 18 and 22 are marked with an ($\times$) as the age was determined from \textit{UBV} alone.} \label{fig:ageVdist}
\end{figure*}

\begin{figure*}[hbt!]
\epsscale{1.05}\plotone{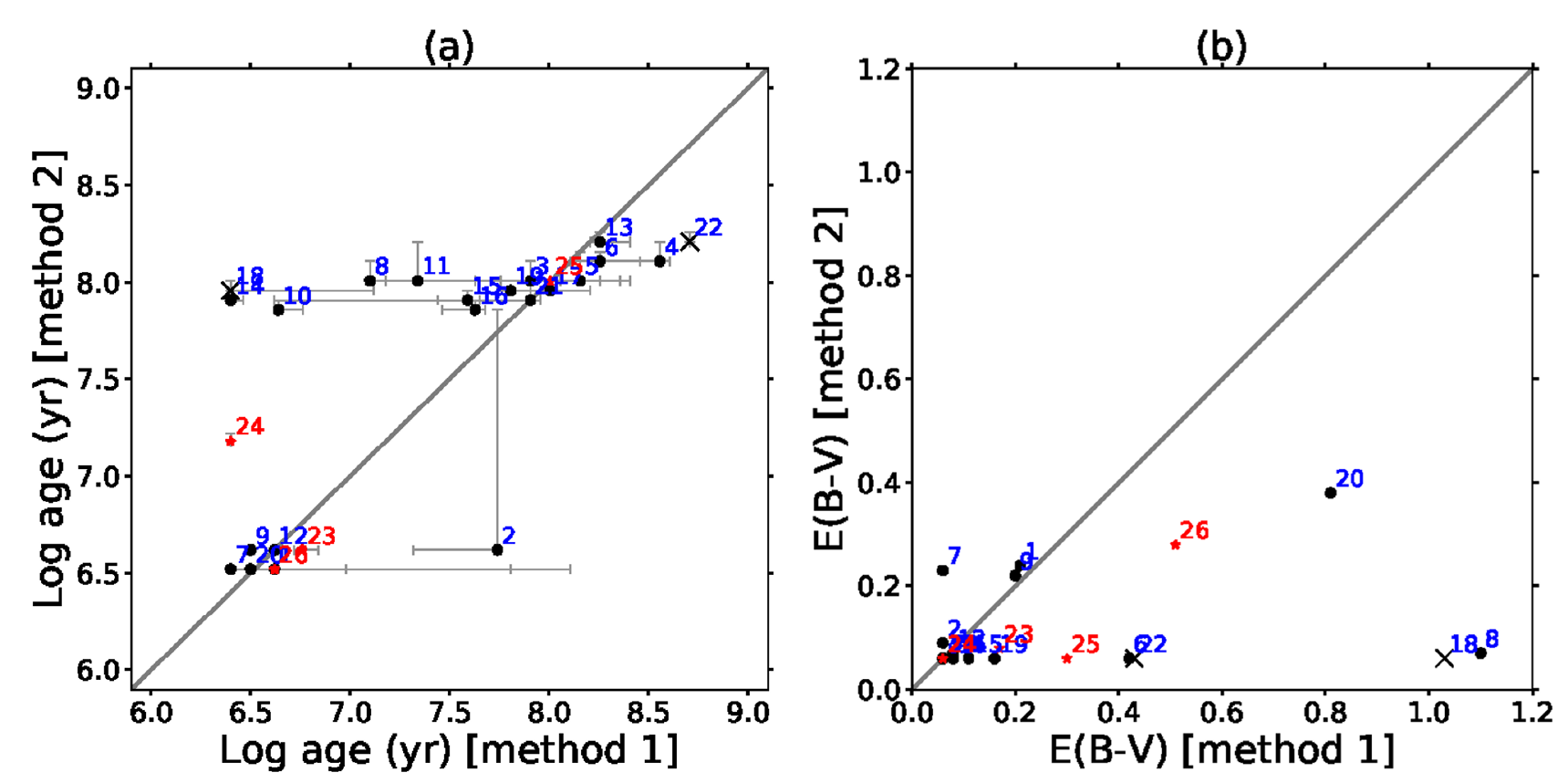}
\caption{Comparison of (\emph{a}) inferred ages, log(age), and (\emph{b}) color excesses, $E(B\!-\!V)$, of the inner regions, computed using either background subtracted fluxes (method 1; horizontal axes) or background subtracted colors (method 2; vertical axes). Age error bars that appear one-sided are the result of the upper or lower color uncertainties finding the same age in the model. The gray line represents where both methods return the same age or $E(B\!-\!V)$. Regions 18 and 22 are marked with an ($\times$) as the age from the background flux subtraction was determined from \textit{UBV} alone. 
The regions without CO (regions 23, 24, 25, and 26) are marked with a red star ($\star$). \label{fig:agemethod}}
\end{figure*}

The projected distance from the CO core to the nearest FUV knot in regions 22 and especially 18 is greater than for most other regions (Figure \ref{fig:ageVdist}) and the NUV or FUV background measured in their annuli exceeded the corresponding average of their inner region. One reason for this may be that the inner star forming region actually extends into the annulus we adopted for the environment. This prevented us from placing meaningful constraints on the \textsl{FUV\,$-$\,NUV} colors of these regions. We see in Figure \ref{fig:ageVdist} that there are possibly two relationships showing the age of a region increasing with distance from its nearest FUV knot (one above log(age) $\simeq$ 7.25 and one below), but there is no clear explanation for why the regions would separate into these two sequences.
As such, only the $U-B$ and $B-V$ colors were used in the age calculation for these two regions. We mark these regions with an ($\times$) when showing the region ages (Figures \ref{fig:ageIn} and \ref{fig:ageAn}). We also examined subtracting the background colors instead of the background fluxes to find the color excess before iterating on reddening, which allowed us to use the \textsl{FUV\,$-$\,NUV}, $U-B$, and $B-V$ colors in the age calculation of all regions. We compare the two methods in Figure \ref{fig:agemethod}, where we see a trade-off between the extinction and the age. Subtracting the background colors typically finds regions to have lower $E(B\!-\!V)$ extinction and higher ages compared to subtracting the background fluxes. We chose to use the ages computed with the background flux subtracted photometry throughout this analysis in order to avoid unphysically old ages for star forming regions. 

To account for the stochastic effects of red supergiants (RSGs), which could skew the colors of small young clusters \citep{Krumholz_2015}, we looked at the location of catalogued RSGs in WLM from \citet{Levesque_2012}. We found three RSGs in four regions (regions 11, 12, 18, and 20), with one RSG located in the overlap of regions 11 and 12. The ages of regions 11, 12, 18, and 20 are rather young at 22, 4, 2, and 3 Myr respectively. With the age of these regions well within the range of all other regions, it is not clear that RSGs have made a noticeable impact on our age calculations.

To find the age of the annulus representing the environment where the star-forming region formed, we did not subtract any background since the age we were measuring was that of the background disk surrounding each region. Likewise, we only computed our measured \textsl{FUV\,$-$\,NUV}, $U-B$, and $B-V$ colors with the $E(B\!-\!V)$ of 0.068 since the annuli are not likely to be heavily reddened. As for the inner regions, our measured colors were fit to the BC03 model colors for each region, where we selected the best fit as the age for that region. Table \ref{tab:datatable3} contains the age of each annulus computed with the Chabrier IMF, and Figure \ref{fig:ageHis} shows a histogram of the annuli ages. 

\subsection{Stellar mass surface density}\label{subsec:massSD}
After finding the age of the inner region using BC03, we took the ratio between the $V$ flux corresponding to the model age of each region and measured the integrated $V$ flux, which was corrected for the extinction we determined from the colors. This ratio was used to scale the model mass and find the mass of young stars ($M_{*,\scriptsize{\textnormal{SF}}}$) for each inner region. 

The young star mass was then divided by the area of each region in parsecs to find the corresponding average $\Sigma_*$ in $M_\odot$ pc$^{-2}$. We note that this method of determining the stellar mass surface density does not take into account the dispersal of stars over time. The model $V$ values corresponding to the upper and lower age uncertainties for each region were used for the model $V$ uncertainties, which were used to compute the uncertainties for the $M_*$ and $\Sigma_*$ for each region. These stellar mass surface densities and uncertainties for the star forming regions can be found in Table \ref{tab:datatable2}. Since the ages were used in calculating the $\Sigma_*$ of the inner regions, regions 18 and 22, where the ages were calculated from \textit{UBV} alone, have their $\Sigma_*$ marked with an ($\times$) in relevant figures.

\subsection{CO-dark H$_2$ gas}\label{subsec:darkgas}
We also wanted to examine any potential relationship of CO-dark H$_2$ gas with the small CO cores. In order to find the percentage of the original molecular cloud mass $M_{MC}$ in CO-dark H$_2$ (\darkgas), we assumed that the percentage of the original total molecular cloud gas that was converted into stars was roughly 2\% \citep{Krumholz_2012}. We used $M_*$ 
and 2\% efficiency to find the total molecular cloud gas mass for each inner region. Then, given that the molecular gas mass of the cloud is a combination of the CO-dark gas mass and the mass contained in CO cores, 
we found the percentage of the molecular cloud that is CO and that is CO-dark. For region 1, we also looked at how much the mass of carbon in the PDR contributes to the mass of the molecular cloud. The [C$\,${\sc ii}]$\lambda$158 $\mu$m flux measurements from \citet{Cigan_2016} correspond to $40-170 M_\odot$ of free carbon atoms. As this is only $0.03-0.14$\% of the original total mass of the molecular cloud, we choose to ignore it in our analysis. We note that the $M_{MC}$ is the original mass of the molecular cloud, which would break up and dissociate with time. Without information on the status of the cloud itself, our estimates do not take into account the current structure of the molecular cloud. 
The uncertainties in the CO-dark H$_2$ mass were found by computing the dark H$_2$ mass percentage using the uncertainties in the $M_*$ for each region as previously described, although the uncertainty is most likely dominated by the assumption of a star formation efficiency. The \darkgas\ and associated uncertainties can be found in Table \ref{tab:datatable2}.
Since the $M_*$ for each region is determined by the $V$ value corresponding to the model age found, regions 18 and 22, for which their age was calculated from \textit{UBV} alone, have their dark H$_2$ mass denoted with an ($\times$) in relevant figures.

In summary, the star forming region properties include the \HI\ surface density (\HISDSF), stellar mass surface density (\MassSDSF), sum of individual CO core virial masses in a region (\MCO), median individual CO core surface density (\COSD) in a region, stellar age, and dark molecular hydrogen to original molecular cloud mass ratio (\darkgas). The environmental properties include the \HI\ surface density (\HISDenv), pressure, stellar mass, and age. In Section \ref{sec:results} we compare and contrast these properties.

\section{Results}\label{sec:results}
\subsection{Characteristics of star-forming regions}\label{subsec:innerEnv}

\begin{figure*}[htb!]
\epsscale{1.17}\plotone{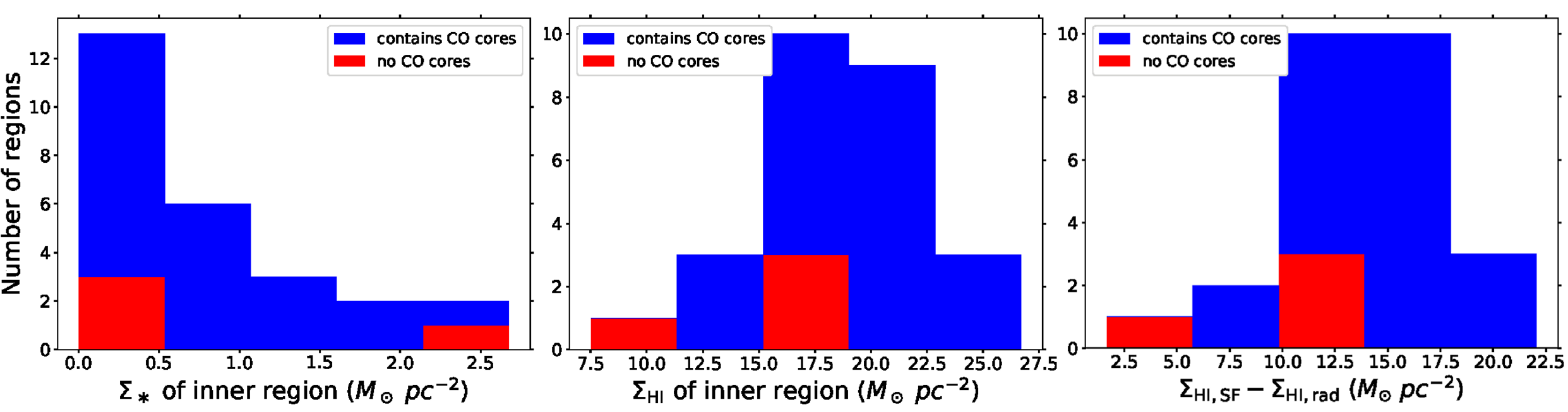}
\caption{Histograms of properties of 
inner regions where CO cores currently sit (blue) and the inner regions without any CO cores (red).
Left: Average stellar mass surface density. Center: Average \HI\ surface density. Right: Difference between the average \HISDSF\ and the corresponding azimuthally-averaged radial profile \HISDrad\ at the region (\HISDdiff).  \label{fig:histIn}}
\end{figure*}

In Figure \ref{fig:histIn} we plot histograms for the star-forming region properties of stellar mass surface density \MassSDSF\ and \HI\ surface density \HISDSF. We also plot a histogram of the difference between the average \HISDSF\ and the corresponding azimuthally-averaged radial profile \HISDrad\ at the region (\HISDdiff).
While the regions without CO cores typically have \MassSDSF, \HISDSF, and \HISDdiff\ values within the range of values for the regions with CO, the \HISDSF\ and \HISDdiff\ tend to fall at lower end of that range. 

\subsubsection{CO core mass and surface density}\label{subsub:COcoresIR}

\begin{figure*}[htb!]
\epsscale{0.7}\plotone{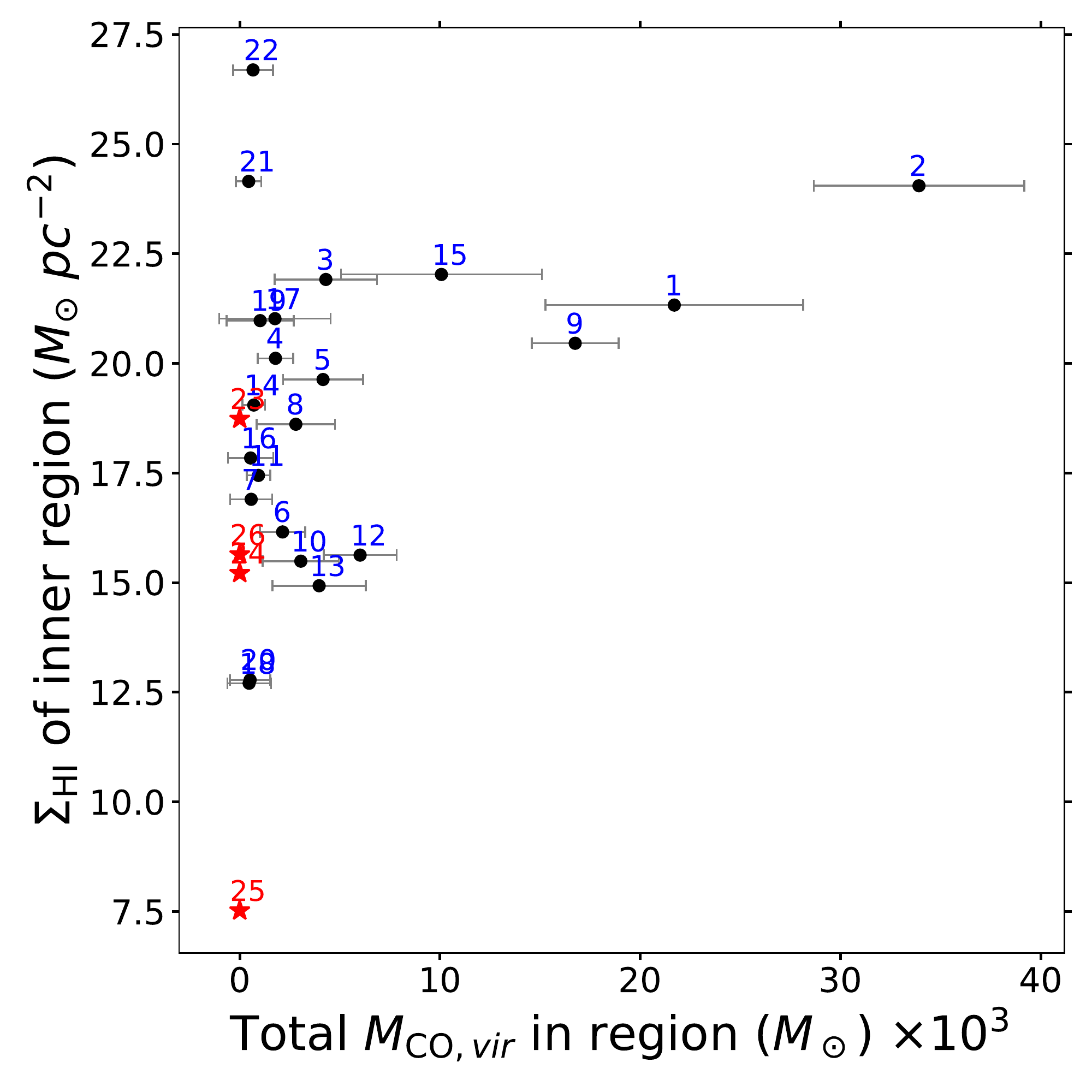}
\caption{\HI\ surface density of each inner region versus the total mass of CO cores of that region, where the \HI\ uncertainties are smaller than the size of the plot markers. We find that a higher \HI\ surface density is found in regions with a higher total CO core mass (regions 1, 2, 9, and 15), while a high \HI\ surface density does not necessarily correspond to a higher total CO core mass (regions 3, 4, 17, 19, 21, and 22).
\label{fig:hiIn}}
\end{figure*}

\begin{figure*}[hbt!]
\epsscale{1.1}\plotone{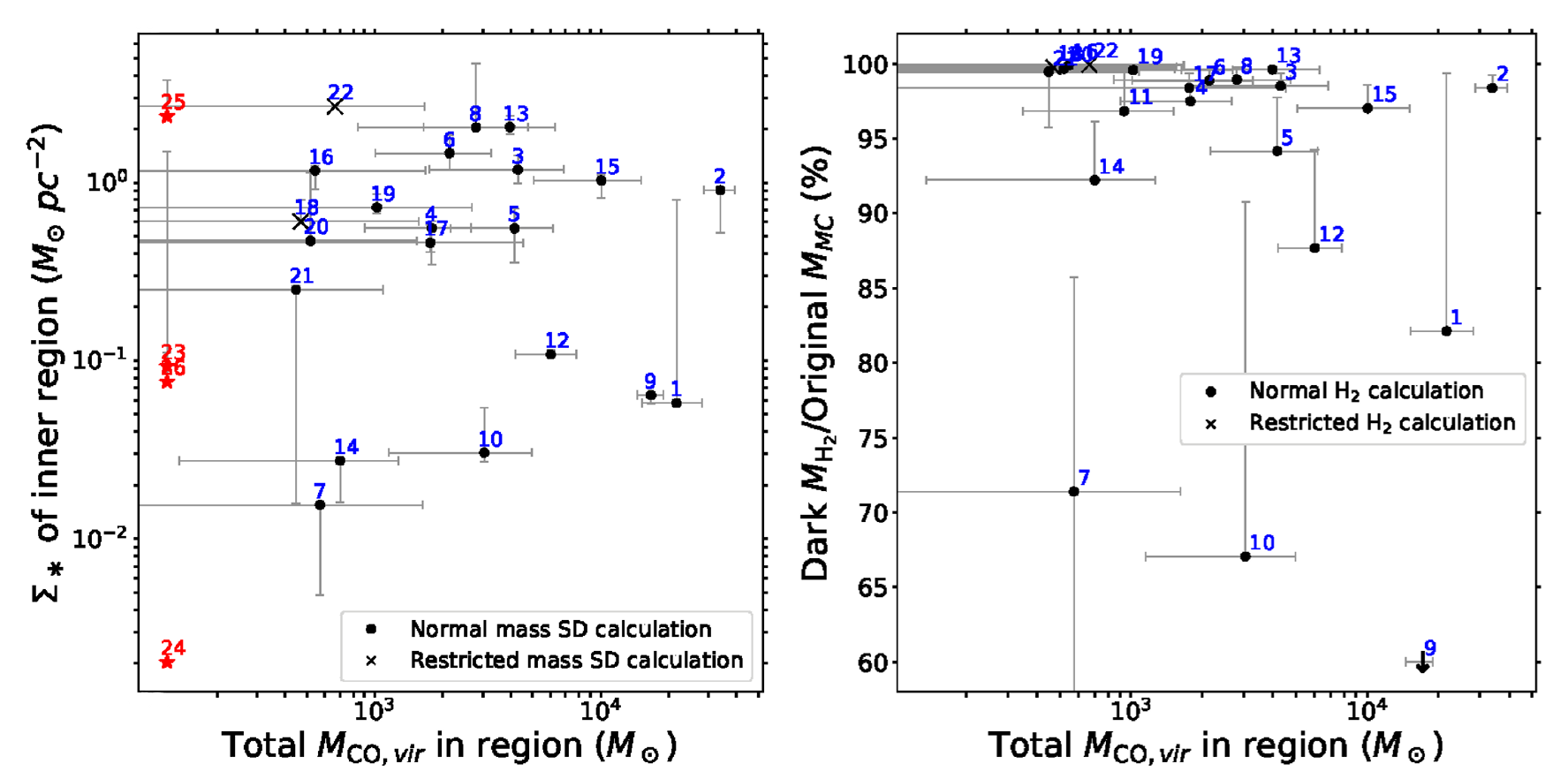}
\caption{Left: Stellar mass surface density of each inner region plotted against the total mass of CO cores of that region. The regions without CO cores (region 23, 24, 25, and 26) are marked with a red star ($\star$) and have a total \MCO\ of zero, but have been placed at a total \MCO\ of $1.2 \times 10^2\ M_\odot$ to show their corresponding $\Sigma_*$. Right: Percentage of original total molecular cloud mass that is dark H$_2$ gas of each inner region plotted against the total mass of CO cores in that region. We do not show the \darkgas\ against the total \MCO\ of regions without detected CO as they all have a total \MCO\ of zero and thus a dark H$_2$ mass percentage of 100\%. Region 9 is displayed with arrow marker to indicate that the actual dark H$_2$ mass ratio value is $0\%$ but has been shifted up to better show the distribution of the dark H$_2$ mass ratios of the other regions. The correct dark H$_2$ mass percentages are listed in Table \ref{tab:datatable2}. Regions 18 and 22 are marked with an ($\times$) as the stellar mass surface density and dark H$_2$ were determined from \textit{UBV} alone.
\label{fig:COmassIn}}
\end{figure*}

\begin{figure*}[hbt!]
\epsscale{1.1}\plotone{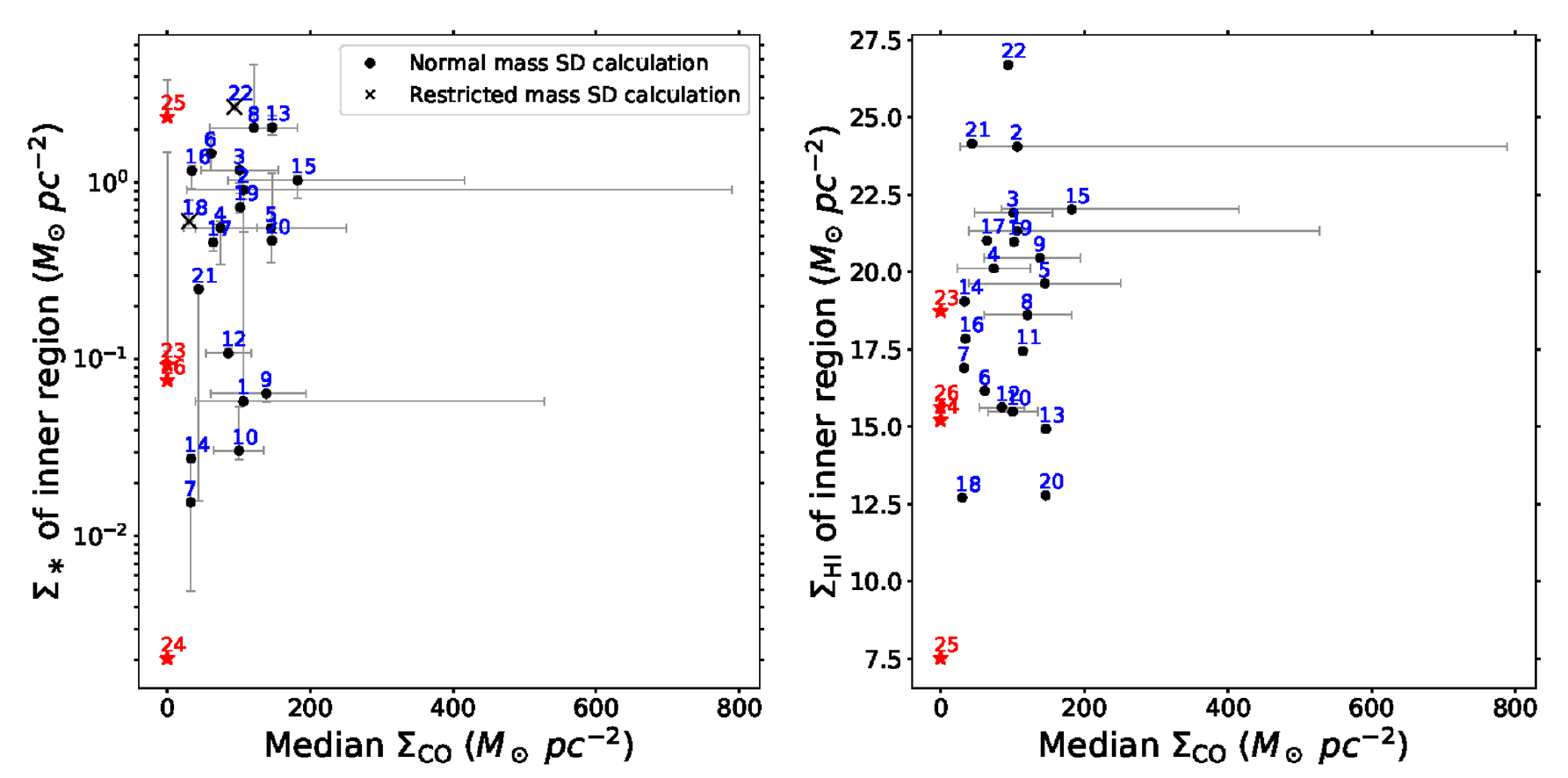}
\caption{Left: Stellar mass surface density of each inner region plotted against the median individual CO core surface density of that region. Regions 18 and 22 are marked with an ($\times$) as the stellar mass surface density was determined from \textit{UBV} alone.
Right: \HI\ surface density of each inner region plotted against the median individual CO core surface density of that region, where the \HI\ uncertainties are smaller than the size of the plot markers. The x-axis error bars are not uncertainties, but instead represent the range of CO core surface densities for that region, with some being smaller than the size of the marker, especially if the region only has one CO core. The regions without CO cores (region 23, 24, 25, and 26) are marked with a red star ($\star$) and have a CO core surface density of zero.
\label{fig:sdIn}}
\end{figure*}

\begin{figure*}[htb!]
\epsscale{1.17}\plotone{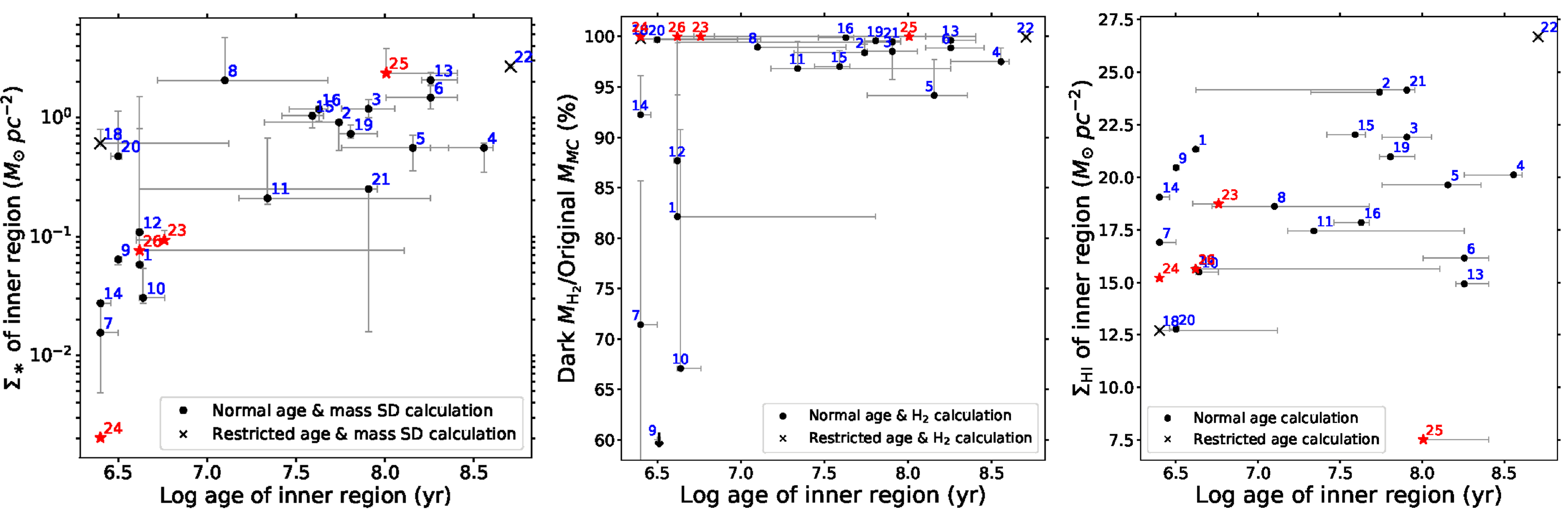}
\caption{Left: Stellar mass surface density of each inner region plotted against the age of that region. Center: ratio (\%) of dark H$_2$ mass to original total molecular cloud mass of each inner region plotted against the age of that region. Region 9 is displayed with an arrow marker to indicate that the actual dark H$_2$ mass ratio value is 0\% but has been shifted up to better show the distribution of the dark H$_2$ mass ratios of the other regions. Right: \HI\ surface density of each inner region plotted against the age of that region, where the \HI\ uncertainties are smaller than the size of the plot markers. Age error bars that appear one-sided are the result of the upper or lower color uncertainties finding the same age in the model. Regions 18 and 22 are marked with an ($\times$) as the age and stellar mass
were determined from \textit{UBV} alone. The regions without CO (region 23, 24, 25, and 26) are marked with a red star ($\star$) and are assumed to have a dark H$_2$ mass percentage of 100\%. Neither the stellar mass surface densities nor dark H$_2$ mass percentages take into account the current structure of the molecular cloud, which would break up and dissociate with time.
\label{fig:ageIn}}
\end{figure*}

Next, we look at whether the sum of individual CO virial masses, \MCO, and median surface density, \COSD, of the individual CO cores in a region have any relationship with the star forming region where they now sit. To do this, we plot the \HISDSF, \MassSDSF, and  \darkgas\ against the total mass of the CO cores in each region. In Figure \ref{fig:hiIn} we see that regions with a higher total \MCO\ also show a higher \HISDSF\ (regions 1, 2, 9, and 15), while a region with a higher \HISDSF\ does not necessarily correspond to a higher total \MCO\ (regions 3, 4, 17, 19, 21, and 22). The correlation between higher \MCO\ and \HISDSF\ suggests that considerable amounts of \HI\ are needed to create large quantities of molecules, and that large molecular clouds are difficult or impossible to make at low \HISD. Figure \ref{fig:COmassIn} shows $\Sigma_*$ and \darkgas\ plotted against the total \MCO\ of each region, where we see no relationship in either.

In Figure \ref{fig:sdIn} we plot \MassSDSF\ and \HISDSF\ with the median individual \COSD\ in each region. Here, the error bars given for the \COSD\ are the minimum and maximum \COSD\ in that region for regions with more than one CO core. We do not find any relation between the \COSD\ in a region and the \MassSDSF\ or \HISDSF\ of that region. 

\subsubsection{Age}\label{subsub:ageIR}
In Figure \ref{fig:ageIn} we plot the \MassSDSF, \darkgas, and \HISDSF\ against the age of the region to examine any relationships between the star forming region properties and the age of the regions where we find CO cores. We find that both the \MassSDSF\ and \darkgas\ appear to increase with age. Both quantities are computed from the age of the region but do not take into account the disruption of the molecular cloud by stars with time, which may affect any apparent correlation seen between the \MassSDSF, \darkgas, and age.

\subsubsection{Dark H$_2$ mass to total molecular cloud mass ratio}\label{subsub:h2IR}
\begin{figure*}[htb!]
\epsscale{1}\plotone{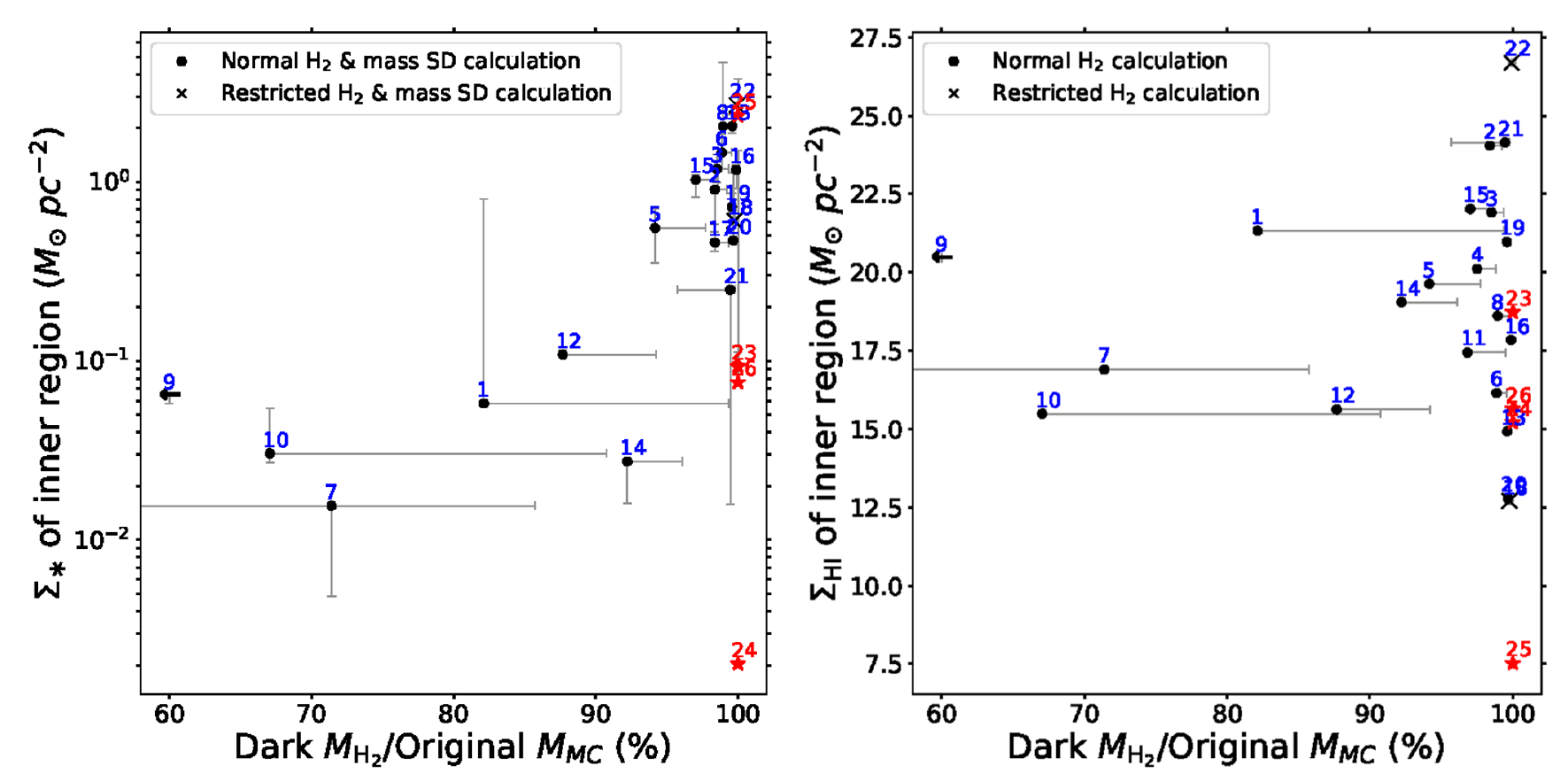}
\caption{Left: Stellar mass surface density of each inner region plotted against the ratio of dark H$_2$ mass to total molecular cloud mass of that region. Right: \HI\ surface density of each inner region plotted against the ratio of dark H$_2$ mass to total molecular cloud mass of that region, where the \HI\ uncertainties are smaller than the size of the plot markers. Region 9 is displayed with a left-pointing arrow marker to indicate that the actual dark H$_2$ mass ratio value is 0\% but has been shifted right to better show the distribution of the other dark H$_2$ mass ratios.
Regions 18 and 22 are marked with an ($\times$) as the stellar mass surface density and dark H$_2$ were determined from \textit{UBV} alone. The regions without CO (region 23, 24, 25, and 26) are marked with a red star ($\star$) and are assumed to have a dark H$_2$ mass percentage of 100\%. Neither the original molecular cloud mass nor the \MassSDSF\ account for the dispersal of the molecular cloud over time, which may affect any correlation seen between the two quantities. The correct stellar mass surface densities and dark H$_2$ mass percentages are listed in Table \ref{tab:datatable2}.
\label{fig:h2In}}
\end{figure*}

In Figure \ref{fig:h2In} we plot the \MassSDSF\ and \HISDSF\ of the inner region against the percentage of the original molecular cloud mass that is in dark H$_2$ to determine if the amount of CO-dark H$_2$ where we find CO cores correlates to other star forming region properties. We find that most regions have a dark H$_2$ mass that is between 90 and 100\% of the original total molecular cloud mass, and that higher \darkgas\ tends to corresponds to higher \MassSDSF. Both of these quantities are derived from the age of the region, while neither take into account dispersal of the molecular cloud over time which may affect any apparent correlation.
The HI surface density of the inner regions varies independently of the percentage of dark gas in the molecular cloud.

\begin{figure*}[htb!]
\epsscale{1.17}\plotone{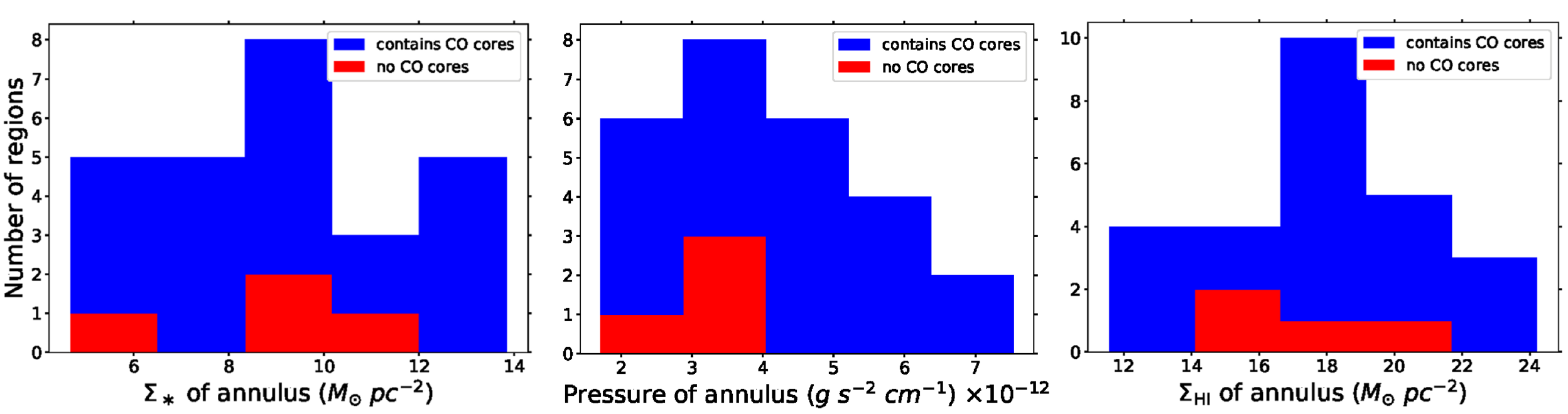}
\caption{Histogram of average stellar mass surface density (left), pressure (center), and \HI\ surface density (right) of annuli representing the environment where CO cores formed (blue) and the annuli of the star forming regions without any CO cores (red). \label{fig:histAn}}
\end{figure*}

\subsection{Environments of star-forming regions}\label{subsec:outerEnv}

To compare the environment where CO cores formed against regions where we do not detect any CO cores, we plot histograms for the environmental properties of \MassSDEnv, pressure, and \HISDenv\ for the outer annulus of all 26 regions. We again find that the annuli surrounding the star forming regions with CO cores fall within the same range of environmental property values as the annuli surrounding regions where no CO cores reside. These histograms are shown in Figure \ref{fig:histAn}.

\subsubsection{CO core mass and surface density}\label{subsub:COcoreAn}

\begin{figure*}[htb!]
\epsscale{1.17}\plotone{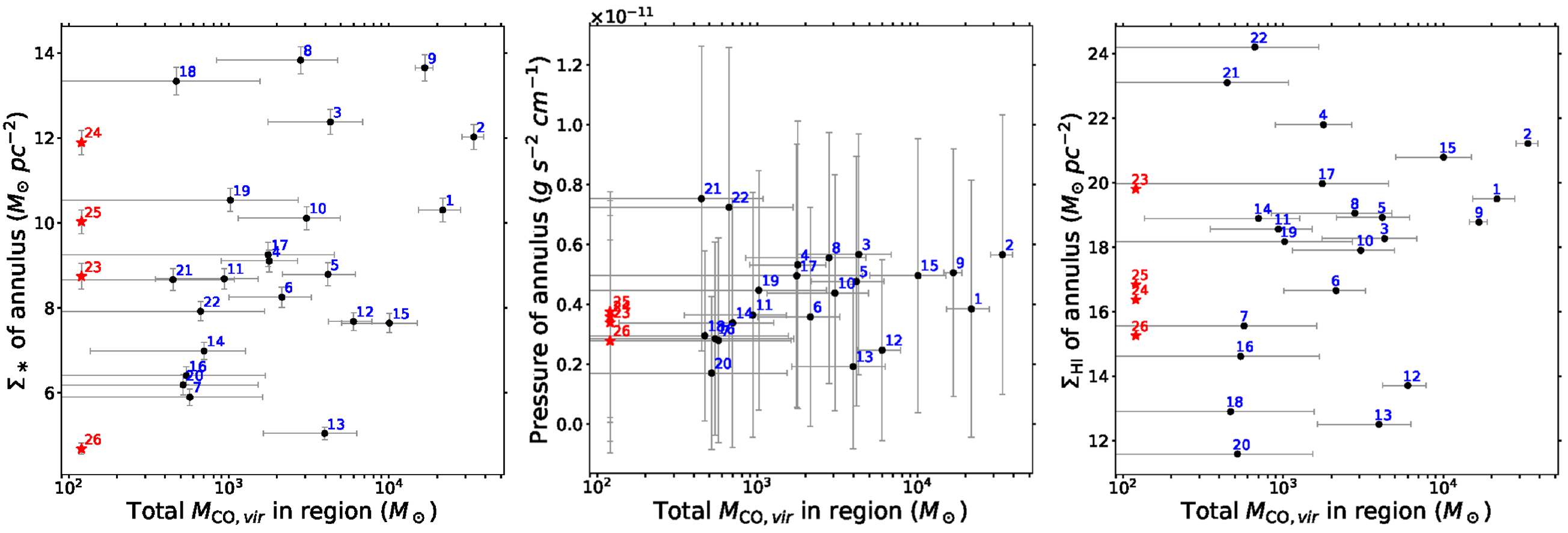}
\caption{Stellar mass surface density (left), pressure (center), and \HI\ surface density (right) of each annulus plotted against the log of the total CO core mass of that region. \HI\ uncertainties are smaller than the size of the plot markers. The regions without CO cores (regions 23, 24, 25, and 26) are marked with a red star ($\star$) and have a total \MCO\ of zero, but have been placed at a \MCO\ of $1.2 \times 10^2\ M_\odot$ to show their corresponding \MassSDEnv, pressure, and \HISDenv.
\label{fig:COmassAn}}
\end{figure*}

\begin{figure*}[htb!]
\epsscale{1.17}\plotone{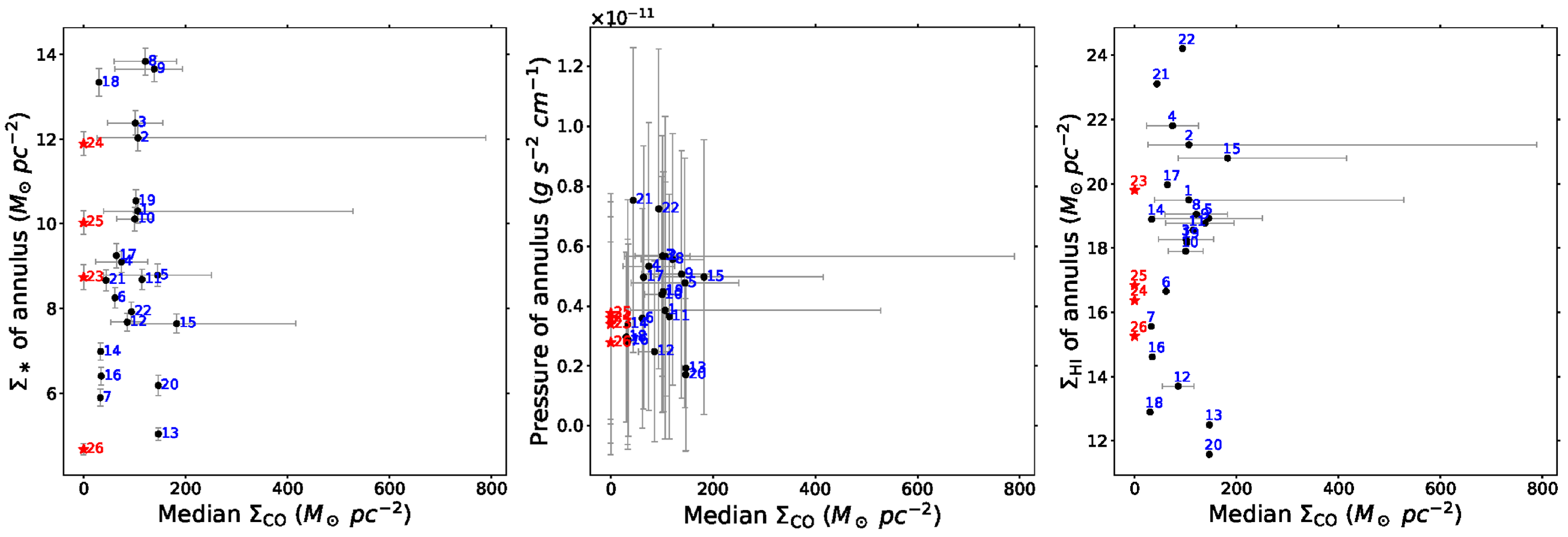}
\caption{Stellar mass surface density (left), pressure (center), and \HI\ surface density (right) of each annulus plotted against the median individual CO core surface density of that region. \HI\ uncertainties are smaller than the size of the plot markers. The x-axis error bars are not uncertainties, but instead represent the range of CO core surface densities for that region, with some being smaller than the size of the marker. The regions without CO cores (regions 23, 24, 25, and 26) are marked with a red star ($\star$) and have a CO core surface density of zero.
\label{fig:sdAn}}
\end{figure*}

\begin{figure*}[htb!]
\epsscale{0.67}\plotone{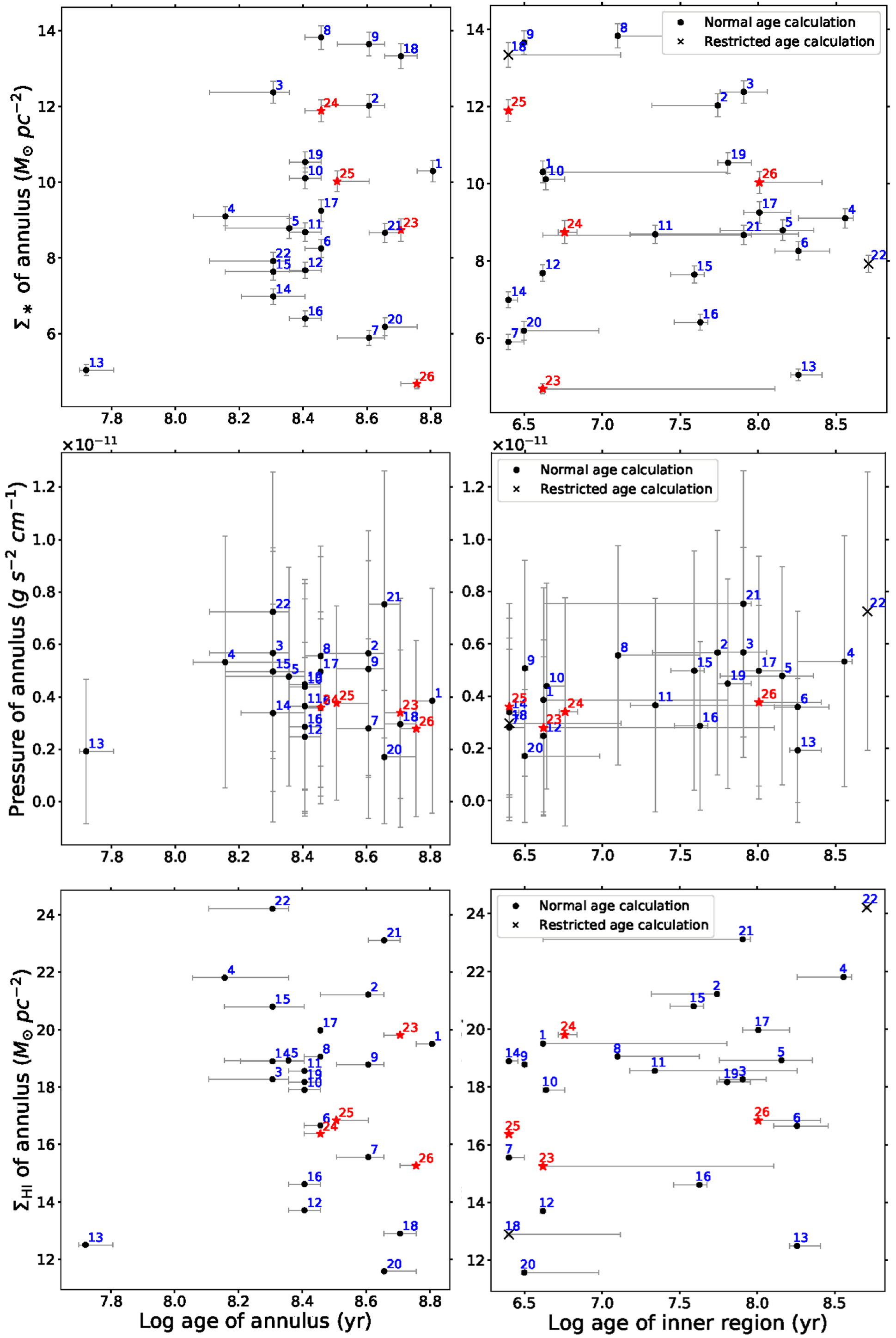}
\caption{Left column: Stellar mass surface density (top), pressure (center), and \HI\ surface density (bottom) of each annulus plotted against the age of that annulus. Right column: Stellar mass surface density (top), pressure (center), and \HI\ surface density (bottom) of each annulus plotted against the age of the corresponding inner region. The age of inner regions spans a much larger range than the age of the annuli, however most star forming regions are less than 100 Myr while the ages of the environment are mostly older than 100 Myr. The \HI\ uncertainties are smaller than the size of the plot markers. Age error bars that appear one-sided are the result of the upper or lower color uncertainties finding the same age in the model. Regions 18 and 22 are marked with an ($\times$) as the age and stellar mass were determined from \textit{UBV} alone. The regions without CO (regions 23, 24, 25, and 26) are marked with a red star ($\star$). \label{fig:ageAn}}
\end{figure*}

Another way of examining the environment where the CO cores formed is to look at the relationship between the sum of individual CO core virial masses \MCO\ of a region with the corresponding \HISDenv, pressure, and \MassSDEnv\ of the annulus of that region.  We show these in Figure \ref{fig:COmassAn}. We find that regions with a higher total \MCO\ tend to have a higher \HISDenv, as we found in the star forming regions themselves, while again showing that a higher \HI\ does not necessarily lead to a higher total \MCO. This correlation between the \HISD\ and total \MCO\ is not as pronounced in the annuli as the inner regions. 
The three regions with highest \MCO\ also have relatively high \MassSDEnv, but
we find no relationship between the total \MCO\ and the pressure.
We also compare the pressure, \HISDenv, and \MassSDEnv\ of the annuli with the median individual \COSD\ of the regions in Figure \ref{fig:sdAn} and find no relationship. 

\subsubsection{Age}\label{subsub:ageAn}

To examine any relationships between the environmental properties and the age of both the environment where the CO cores formed and the star forming region where we now find the CO cores, we plot the environmental pressure, \HISDenv, and \MassSDEnv\ against the age of the annuli and the age of the inner region in Figure \ref{fig:ageAn}. Not surprisingly we find that the age of the annuli are older than inner regions. The annuli ages of the regions fall between $\sim$50 and 650 Myr, with most around 250 Myr. The age of inner regions spans a much larger range between $\sim$2 and 500 Myr, with most regions less than 100 My. We do not find any correlation between the environment where the CO cores formed and either the current age of that environment or the age of the star-forming region in which the CO cores sit. 

\subsubsection{Dark H$_2$ mass to original total molecular cloud mass ratio}\label{subsub:h2An}

\begin{figure*}[htb!]
\epsscale{1.15}\plotone{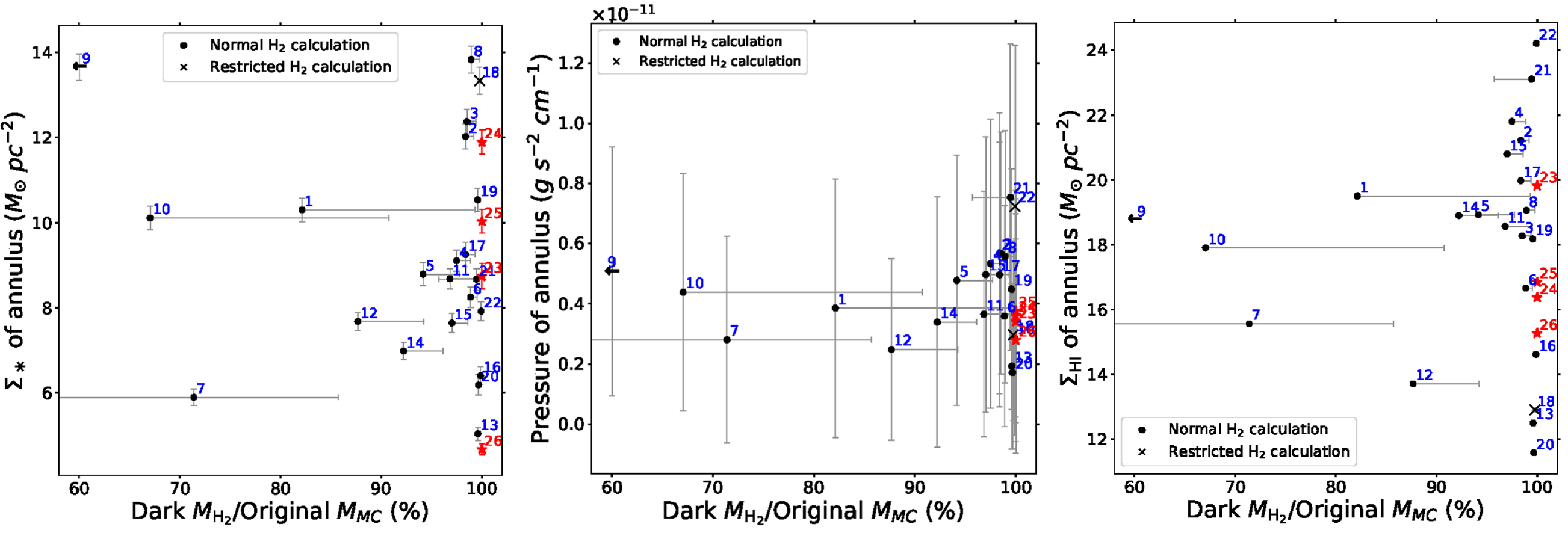}
\caption{Stellar mass surface density (left), pressure (center), and \HI\ surface density of each annulus plotted against the dark H$_2$ mass to original total molecular cloud mass ratio of that region. \HI\ uncertainties are smaller than the size of the plot markers. Region 9 is displayed with a left-pointing arrow marker to indicate that the actual dark H$_2$ mass ratio is 0\% but has been shifted right to better show the distribution of the other dark H$_2$ mass ratios. The total molecular cloud mass is the original mass of the cloud and does not take into account the dissociation of the cloud over time. 
Regions 18 and 22 are marked with an ($\times$) as the dark H$_2$ were determined from the stellar mass that was derived from \textit{UBV} alone. 
The regions without CO (regions 23, 24, 25, and 26) are marked with a red star ($\star$) and assumed to have a dark H$_2$ mass percentage of 100\%. \label{fig:h2An}}
\end{figure*}

In Figure \ref{fig:h2An} we plot the \MassSDEnv, pressure, and \HISDenv\ of the annulus of each region against the \darkgas\ for that region to determine if the amount of CO-dark H$_2$ in the star forming regions where the CO cores sit have any relationship with the environmental properties where the CO cores formed. We find no correlations between the percentage of original molecular cloud mass in dark H$_2$ and the \MassSDEnv, pressure, or \HISDenv\ of the annuli where the CO cores formed.

\subsection{Summary of results}\label{sec:summary}
\citet{Rubio:2015} and Rubio et al.\ (2022, in preparation) discovered CO cores in the dIrr galaxy WLM, which has a metallicity 13\% of solar. The detection of this CO is important for understanding star formation in the most numerous type of galaxy, as CO is used to trace the molecular hydrogen thought to be responsible for star formation. This study is aimed at understanding the environments in which these small CO cores form at low mass and metallicities. In this work, we have examined the properties of CO-detected regions in WLM and explored relationships between the CO and the environments where they formed and the star-forming regions where they currently reside. We grouped the cores into 22 regions based on proximity to FUV knots, along with four regions containing FUV emission that don't have any detected CO cores. 

We looked at the \MassSDSF, \HISDSF, total \MCO, median individual \COSD, age, and \darkgas\ of the star forming regions and the \MassSDEnv, pressure, \HISDenv, and age of the environments measured in annuli around the star-forming regions. 
We do not see any difference between the star forming region properties where we find CO cores and the star forming region properties where we do not find CO cores, nor do we see a difference between environmental region properties where we find CO cores and environmental region properties where we do not. We do not see any correlations among the star forming region properties or environmental region properties except between the \HISD\ and total \MCO\ and, to a lesser extent, the pressure and total \MCO. We find that regions with a higher total \MCO\ have higher \HI\ surface densities, and this relationship is more pronounced in the star forming regions than in the surrounding environment.

\begin{figure*}[htb!]
\epsscale{1}\plotone{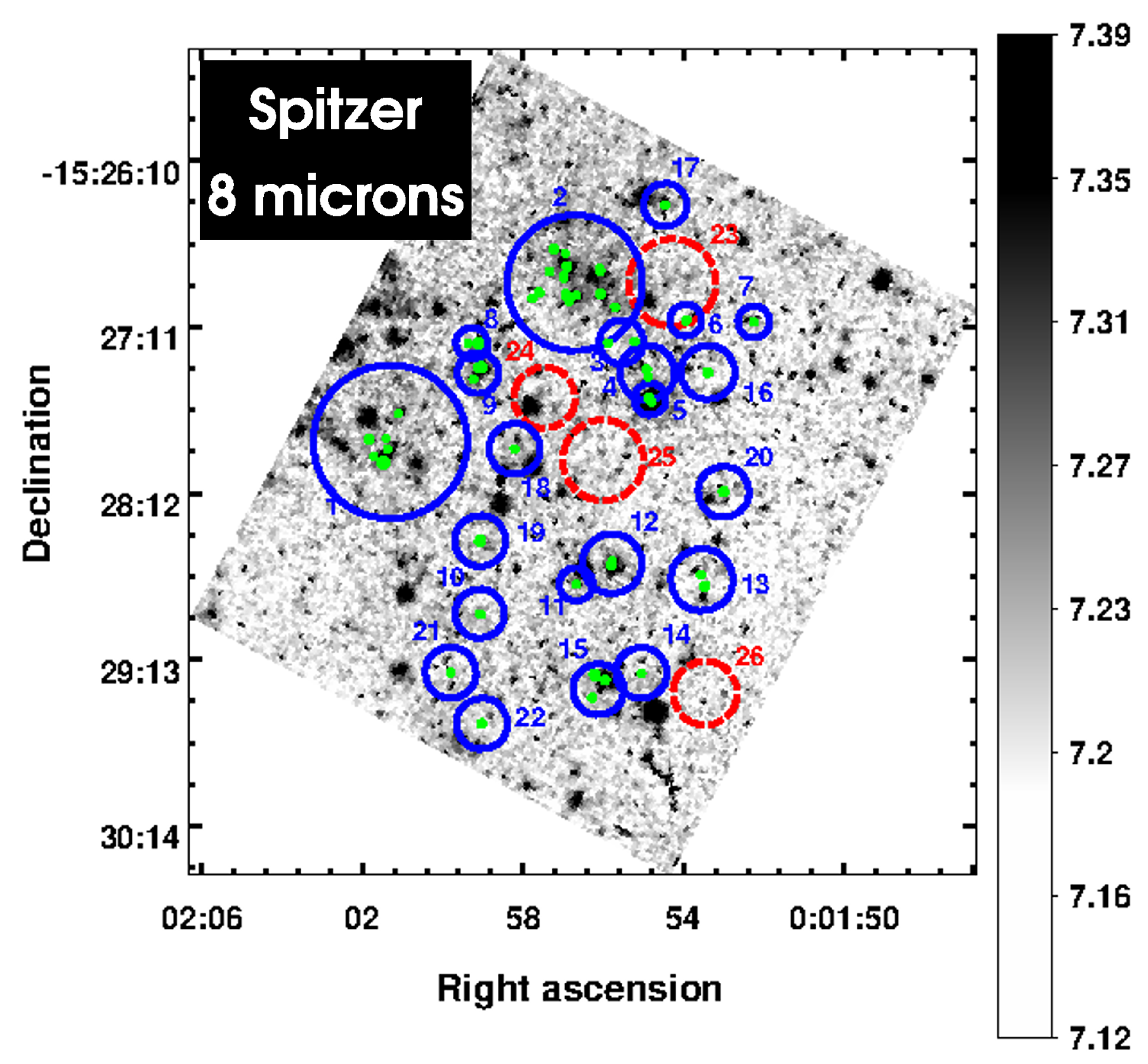}
\caption{Spitzer 8 $\mu$m image of WLM showing the regions defined here (larger blue and red dashed circles) and the CO cores (tiny green circles). Colorbar values are in units of MJy sr$^{-1}$. \label{fig:spitzer}}
\end{figure*}

\section{Discussion and Conclusions}\label{sec:discussion}
Regions 1, 2, 9, and 15 have the highest number of CO cores (6, 17, 4, and 3 cores respectively) and, as expected, the highest total \MCO\ of the regions. We calculate the amount of dark H$_2$ in a region using the assumption that 2\% of the total molecular gas (dark H$_2$ plus \MCO) is turned into stars, and find that the percentage of CO-dark H$_2$ of regions 1, 2, and 15 agree with that of the other regions. Region 9, however, has a \darkgas\ of 0\%. The total \MCO\ of the region is higher than what the total molecular cloud gas mass is expected to be with a 2\% star formation efficiency. One possible reason for this is embedded star formation. To look for potential embedded star formation, we show the regions and their CO cores overlaid on the Spitzer 8 $\mu$m image of WLM in Figure \ref{fig:spitzer}. Here we see 8 $\mu$m peaks near several CO cores, including those in region 9. This may suggest that region 9 has yet to convert 2\% of its molecular gas to stars.

\begin{figure*}[htb!]
\hspace*{-1.0cm}
\epsscale{0.8}\plotone{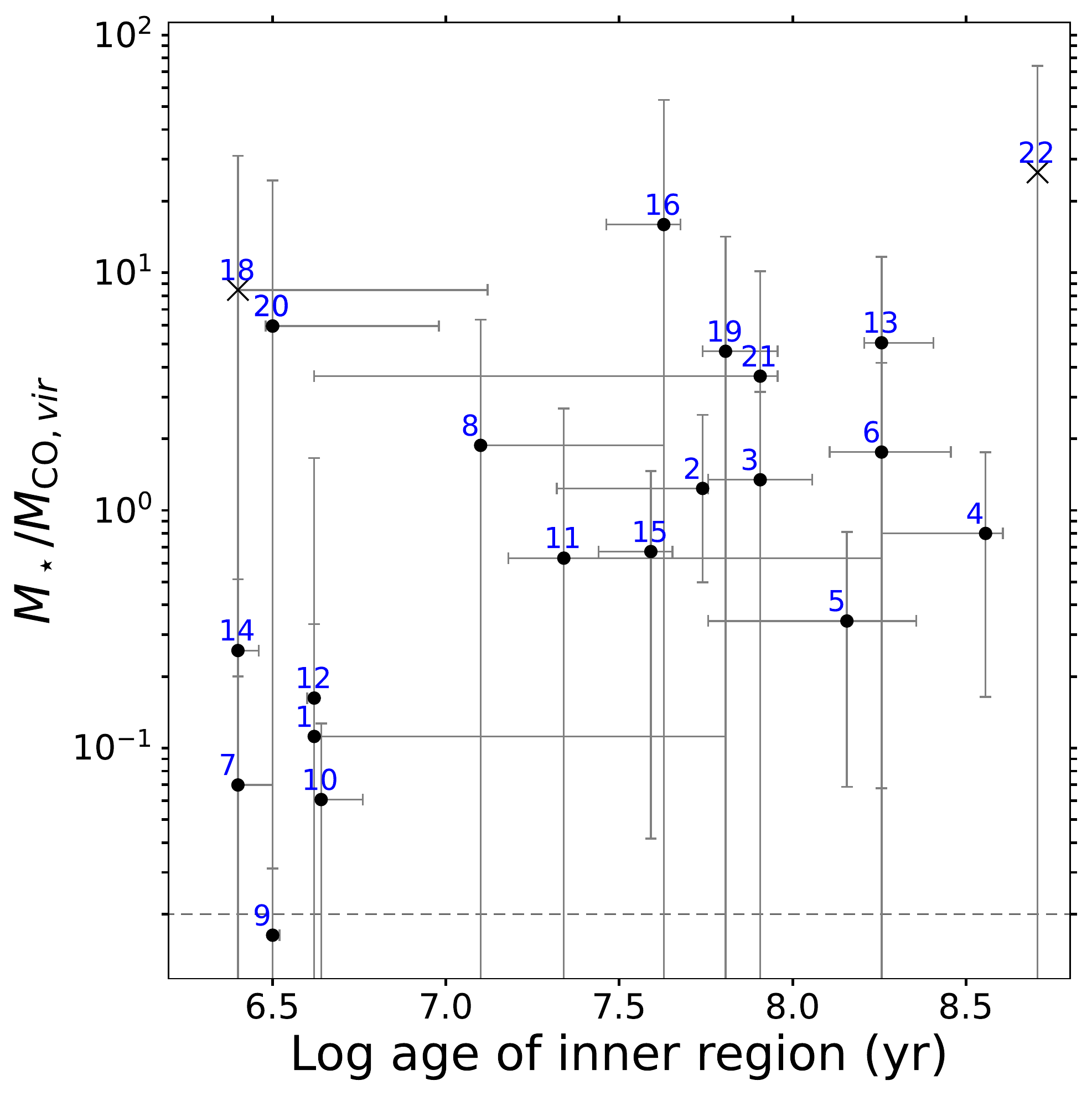}
\caption{Ratio of young stellar mass to observed CO virial mass in the inner region ($M_\ast$/\MCO) against the age of the inner region. The gray dashed line at $M_\ast$/\MCO$=2\%$ shows where $M_\ast$/\MCO\ is equal to the star formation efficiency of 2\%. \label{fig:mStarCOratio}}
\end{figure*}

In Figure \ref{fig:mStarCOratio} we plot the ratio of the young stellar mass to the observed CO virial mass ($M_\ast$/\MCO) of the inner regions against the age of the inner regions. Because we made the assumption that $M_\ast = 0.02 (M_{CO,vir} + M_{Dark\ H_2})$, we mark with a gray dashed line where the $M_\ast$/\MCO\ is 2\% to examine what the excess mass above our estimated star formation efficiency is in dark CO. For young regions (1, 7, 10, 11, 14) we find that the excess of $M_\ast$/\MCO\ above 2\% is 
\begin{equation}
\left(1+\frac{M_{Dark\ H_2}}{M_{CO,vir}}\right)\approx5,
\end{equation}
which yields
\begin{equation}
\frac{M_{Dark\ H_2}}{M_{CO,vir}}\approx 4.
\end{equation}
This value agrees with that found in larger scale regions in recent papers \citep{Hunter_2019,Hunter_2021}. When the ratio $M_\ast$/\MCO\ is much larger than $\sim$4, as we see it is for mostly old regions, the molecular gas has likely been destroyed. We see this transition at an age of about 5 Myr which is reasonable for the time it takes for young star formation to break apart its GMC \citep{Williams_2000,Kim_2018,Kruijssen_2019}.

Looking at the scale of the whole galaxy, we see in Figure \ref{fig:hiWreg} a ridge or shell surrounding a depression of \HI. Star formation, shown in Figure \ref{fig:fuvWreg},
is found within and along the ridge as well as further into the hole.
A possible scenario is that past star formation within the hole pushed the \HI\ gas outward and created the ridge we see today \citep{Heiles_1979, Meaburn_1980, Hunter_2001,Kepley_2007}. However, we would then expect to see an age gradient, with the oldest regions closer to the center of the hole, but in fact we do not see any systematic pattern of ages.

The wide range of ages for our star forming regions and their lack of correlation with total \MCO\ also suggests that the extremely small size of the individual CO cores is not due only to fragmentation of aging clouds. Instead, tiny CO cores are all that can be formed in a galaxy with this gas density and metallicity without some galaxy-scale compression.

\citet{Hunter_2001} find star formation is located where \HISD\ is locally higher in the dIrr galaxy NGC 2366 which, like WLM, has a ring of \HI\ surrounding most of the star formation in the galaxy. In Figure \ref{fig:histIn} we see that star forming regions with CO cores have an average \HISD\ higher than the radial average by amounts of 8-22 $M_\odot$, which is consistent with the need for a higher \HISD\ than the average in dIrrs to form stars. This, along with the relationship between higher \HISD\ and higher total \MCO\ suggests that \HISD\ may play a role in the formation of these CO cores. However, the presence of star forming regions with lower \MCO\ along this \HI\ ridge suggests that a higher \HISD\ does not guarantee their formation. We also find star forming regions with CO cores that are not on this high density \HI\ ridge,
which we would not expect to see if higher \HISD\ or pressure were needed to form CO cores. This could mean that the ridge of \HI\ that we see is actually a ``bubble" that we are only seeing in two dimensions. Additionally, there are portions of the \HI\ ridge without any star formation associated with it, particularly to the southeast, that do not show any obvious difference from the rest of the ridge. This portion of the ridge may contain CO cores, but was not surveyed due to lack of time. However, the area surveyed is still representative of the star forming area of the galaxy. The presence of CO cores in a variety of different local environments, along with the similar properties between star forming regions containing CO cores and those without CO cores, leads us to conclude that we do not find clear characteristics to form star forming regions with CO cores.

\acknowledgments
We are grateful for funding for this work, which was provided by the National Science Foundation through grant AST-1907492 to D.H. M.R. wishes to acknowledge support from ANID(CHILE) through FONDECYT grant No1190684.

This paper makes use of the following ALMA data: ADS/JAO.ALMA\texttt{\#}2012.1.00208.S and ADS/JAO.ALMA\texttt{\#}2018.1.00337.S. ALMA is a partnership of ESO (representing its member states), NSF (USA) and NINS (Japan), together with NRC (Canada), MOST and ASIAA (Taiwan), and KASI (Republic of Korea), in cooperation with the Republic of Chile. The Joint ALMA Observatory is operated by ESO, AUI/NRAO and NAOJ. The National Radio Astronomy Observatory is a facility of the National Science Foundation operated under cooperative agreement by Associated Universities, Inc.

Lowell Observatory sits at the base of mountains sacred to tribes throughout the region. We honor their past, present, and future generations, who have lived here for millennia and will forever call this place home.

\vspace{5mm}
\facilities{ALMA, VLA, GALEX, Lowell Observatory:1.1m}
\software{IRAF \citep{Tody_1986}, SExtractor \citep{Bertin_1996}}

\bibliography{2021WLM}{}
\bibliographystyle{aasjournal}



\end{document}